\documentclass{amsart}
\pdfoutput=1
\usepackage[T1]{fontenc}
\usepackage[utf8]{inputenc}
\usepackage[english]{babel}

\usepackage{natbib}

\usepackage{enumerate}

\usepackage{tikz}
\usetikzlibrary{matrix}
\usetikzlibrary{positioning}
\usetikzlibrary{arrows}
\usetikzlibrary{chains}
\usetikzlibrary{calc}
\usetikzlibrary{shapes.geometric}
\usetikzlibrary{fit}

\newtheorem{thm}{Theorem}
\newtheorem{prop}[thm]{Proposition}
\newtheorem{lem}[thm]{Lemma}

\theoremstyle{definition}
\newtheorem*{exmp}{Example}
\newtheorem{defn}{Definition}
\theoremstyle{remark}
\newtheorem*{rem}{Remark}
\newtheorem*{ack}{Acknowledgements}

\newenvironment{pf}{\begin{proof}}{\end{proof}}

\makeatletter
\newif\if@restonecol
\makeatother
\usepackage[algo2e,vlined,linesnumbered,ruled]{algorithm2e}
\dontprintsemicolon
\SetKwSty{}
\SetCommentSty{textsl}
\SetFuncSty{textrm}
\SetKwFor{For}{for}{do}{}
\SetKwComment{Rem}{$\triangleright$~}{}
\SetKwFor{Function}{function}{}{end}
\SetKwFunction{Asympt}{Asympt}
\SetKwFunction{RecToDiffeq}{RecToDiffeq}
\SetKwFunction{NormalizedDiffeq}{Normalize}
\SetKwFunction{BoundRatpoly}{BoundRatpoly}
\SetKwFunction{BoundNormalDiffeq}{BoundNormalDiffeq}
\SetKwFunction{BoundRec}{BoundRec}
\SetKwFunction{HumanReadableBound}{HumanReadableBound}

\DeclareUnicodeCharacter{00A0}{\nobreakspace}
\DeclareUnicodeCharacter{00B2}{\ensuremath{^2}}
\DeclareUnicodeCharacter{00AB}{<<}
\DeclareUnicodeCharacter{00BB}{>>}
\DeclareUnicodeCharacter{00B3}{\ensuremath{^3}}
\DeclareUnicodeCharacter{00D7}{\ensuremath{\times}}
\DeclareUnicodeCharacter{00B1}{\ensuremath{\pm}}
\DeclareUnicodeCharacter{00B0}{\textdegree}
\DeclareUnicodeCharacter{00B7}{\cdot}

\newcommand\N{\mathbb N}
\newcommand\Z{\mathbb Z}
\newcommand\R{\mathbb R}
\newcommand\CC{\mathbb C}

\newcommand\Q{\mathbb Q}
\newcommand\maj\trianglelefteq
\newcommand{\abs}[1]{\left|#1\right|}
\newcommand{\floor}[1]{\left\lfloor#1\right\rfloor}
\newcommand{\ceil}[1]{\left\lceil#1\right\rceil}
\newcommand{\rising}[1]{\overline{#1}}

\DeclareMathOperator\ord{ord}

\DeclareMathOperator\den{den}

\DeclareMathOperator\Ai{Ai}
\DeclareMathOperator\Bi{Bi}
\DeclareMathOperator\Sinei{Si}
\DeclareMathOperator\erf{erf}
\DeclareMathOperator\arccot{arccot}

\usepackage[pdftex,bookmarks,bookmarksnumbered,letterpaper,pdfborder={0 0 0}]{hyperref}

\hypersetup{
pdftitle = {Effective Bounds for P-Recursive Sequences},
pdfauthor = {Marc Mezzarobba, Bruno Salvy},
pdflang = {en},
pdfkeywords = {algorithm, bounds, Cauchy-Kovalevskaya majorant, certified evaluation, holonomic functions}
}

\begin{document}

\title{Effective Bounds for P-Recursive Sequences}

\thanks{This research has been supported in part by the joint Inria-Microsoft research laboratory.}

\author{Marc Mezzarobba}
\email{Marc.Mezzarobba@inria.fr}
\urladdr{http://algo.inria.fr/mezzarobba}
\author{Bruno Salvy}
\email{Bruno.Salvy@inria.fr}
\urladdr{http://algo.inria.fr/salvy}

\address{Algorithms Project, Inria Paris-Rocquencourt, France}

\begin{abstract}
We describe an algorithm that takes as input a complex sequence $(u_n)$ given by a linear recurrence relation with polynomial coefficients along with initial values, and outputs a simple explicit upper bound $(v_n)$ such that $\abs{u_n} \leq v_n$ for all $n$. Generically, the bound is tight, in the sense that its asymptotic behaviour matches that of $u_n$. 
We discuss applications to the evaluation of power series with guaranteed precision.
\end{abstract}


\maketitle


\section{Introduction}

A sequence $u \in \CC^\N$ is \emph{polynomially recursive}, or \emph{P-recursive} (over $\Q$) if it satisfies a non-trivial linear recurrence relation
\begin{equation} \label{eq:prec}
p^{[s]}(n) \, u_{n+s} + \cdots + p^{[1]}(n) \, u_{n+1}  + p^{[0]}(n) \, u_{n} = 0 
\end{equation}
with polynomial coefficients $p^{[k]} \in \Q[n]$. Likewise, an analytic function (or a formal power series) $u$ is \emph{differentially finite}, or \emph{D-finite}, if it is solution to a non-trivial linear differential equation
\begin{equation} \label{eq:dfinite}
p^{[r]}(z) \, u^{(r)}(z) + \cdots + p^{[1]}(z) \, u'(z)  + p^{[0]}(z) \, u(z) = 0, \qquad p^{[k]} \in \Q[z]. 
\end{equation}
The coefficients of a D-finite power series form a P-recursive sequence, and conversely, the generating series of a P-recursive sequence is D-finite. Numerous sequences arising in combinatorics are P-recursive, while many elementary and special functions are D-finite.

Starting with the works of~\citet{Stanley1980}, \citet{Lipshitz1989} and \citet{Zeilberger1990}, D-finiteness relations have gradually been recognized as good \emph{data structures} for symbolic computation with these analytic objects. This means that many operations of interest may be performed on the implicit representation of sequences and functions provided by an equation such as \eqref{eq:prec}, \eqref{eq:dfinite} along with sufficiently many initial values \citep[see][]{SalZim1994, Stanley1999}. In recent years, significant research efforts have been aimed at developing and improving algorithms operating on this data structure.

In this article, we describe an algorithm for computing upper bounds on P-recursive sequences of complex numbers. Specifically, we prove the following theorem (whose vocabulary is made more precise in the sequel).

\begin{thm} \label{thm:tightness}
Given as input a reversible recurrence relation of the form \eqref{eq:prec} with rational coefficients along with initial values defining a sequence $(u_n) \in \Q[i]^\N$, Algorithm~\ref{algo:boundrecdoit} computes $A \in \R_+$, $\kappa \in \Q$, $\alpha \in \bar \Q_+^\ast$ (the set of positive algebraic numbers) and $\phi$ such that
\begin{equation} \label{eq:thm}
\forall n \in \N, \quad \abs{u_n} \leq A \, n!^\kappa \, \alpha^n \, \phi(n); 
\end{equation}
with $\phi(n) = e^{o(n)}$. Moreover, for generic initial values, $\kappa$ and $\alpha$ are tight.
\end{thm}

Asymptotic expansions of P-recursive sequences are a well-studied subject \citep[see, e.g.,][]{Odlyzk1995, AnalyticCombinatorics} and their computation has been largely automated \citep{WimZei1985, Tournier1987, FlSaZi1991, Zeilbe2008}.
While an asymptotic estimate gives a precise indication on the behaviour of the sequence for large values of its index, it cannot in general be used to get an estimate for a specific value. Our result lets one obtain explicit bounds valid for any term, while the tightness of the bound with respect to the asymptotic behaviour implies that the bound is not straying too far away from the actual value. These bounds may be useful both inside rigorous numerical algorithms for problems such as D-finite function evaluation or numerical integration, or as ``standalone'' results to be reported to the user of a computer algebra system. The problem of accuracy control in several settings covering the evaluation of D-finite functions has been considered by many authors~\citep[see in particular][]{Hoefke2001, MakBer2003, Neher2003, Rihm1994a, vdH2003, vdH2007b}. We review previous work on this problem in some more detail in~§\ref{sec:num}. Our main contribution from this viewpoint is to give bounds that are asymptotically tight.

\begin{exmp} \label{ex:intro}
To get a sense of the kind of bounds we can compute, consider the following examples. For readability, the constants appearing in the polynomial parts of the bounds are replaced by low-precision approximations.
\begin{enumerate}[(a)]
\item \label{item:WZ1} Suppose we want to bound
\[ I_n = \int_0^\infty t^n e^{-t^2-1/t} \, dt \]
as a function of $n\in\N$. From the recurrence relation $2I_{n+3}=(n+2)I_{n+1}+I_n$ and the initial conditions $I_0, I_1, I_2 \leq 1/5$, Algorithm~\ref{algo:boundrecdoit} finds that
\[ I_n \leq n!^{1/2} 2^{-n/2} \cdot (0.26\,n+0.76) \binom{n+19}{19}. \]
In fact, $I_n \sim n!^{1/2} 2^{-n/2 - 3/4} (\pi/n)^{3/4}$ as $n\rightarrow\infty$, so that with the notations of Theorem~\ref{thm:tightness}, $\kappa=1/2$, $\alpha=2^{-1/2}$ are indeed recovered by our algorithm.
(This example and the following one are adapted from \citet[Examples~2.1 and 2.3]{WimZei1985}, who illustrate the computation of asymptotic expansions by the Birkhoff-Trjitzinsky method.)
\item \label{item:WZ2} The number $t_n$ of involutions of $\{1, \dots, n\}$ satisfies the recurrence relation
\[ t(n+2)=(n+1)t(n)+t(n+1), \qquad t(0) = t(1) = 1, \] 
and $t_n \sim (8\pi)^{-1/4} n!^{1/2} e^{\sqrt{n}-1/4} n^{-1/4}$ as $n \to \infty$ \citep[see][§5.1.4]{TAoCP2}. Assume that we wish to bound the probability that a permutation chosen uniformly at random is an involution: the same algorithm leads to\footnote{We use $[z^n]f$ to denote the coefficient of $z^n$ in the power series~$f$, see the end of~\S1 for notations.}
\[ \frac{t(n)}{n!} \leq (0.90\,n+2.69) \, n!^{-1/2} \, [z^n]\exp \frac{1}{1-z} = O( n^{1/4} \, n!^{-1/2} \, e^{2 \sqrt{n}} ). \]
Compare \citep[Example~VIII.5]{AnalyticCombinatorics}. Notice that, in addition to the parameters $\alpha$ and $\kappa$ of Theorem~\ref{thm:tightness}, the subexponential growth type $e^{O(\sqrt n)}$ is preserved. However, our algorithm is not designed to preserve the constant in this $O(\cdot)$ term.
\item \label{item:exchudnovsky} One of the fastest ways to compute high-precision approximations of $\pi$ resorts to the following formula due to~\citet[p.~389]{ChuChu1988}:
\[
\sum_{k=0}^\infty t_k = \frac{640320^{3/2}}{12\pi}
\qquad \text{where} \quad
t_k = \frac{(-1)^k (6k)! (13591409 + 545140134k)}{(3k)!(k!)^3 640320^{3k}}.
\]
Using the method of §\ref{sec:tails} on the obvious first order recurrence relation satisfied by $(t_k)$, our algorithm leads to 
\[ \left| \sum_{k=n}^{\infty} t_k \right|
\leq 10^6 (2.3\,n^3+ 13.6\,n^2 + 25\,n + 13.6) \alpha^n \]
where $\alpha = \frac{1}{151931373056000} \simeq 0.66 \cdot 10^{-14}$. We see that each term of the series gives about $14$ more correct decimal digits of $\pi$, and we can easily deduce a suitable truncation order to compute $\pi$ to any given precision.
\item \label{item:Si} Similarly, from the differential equation
\[
z\,\Sinei'''(z) + 2\,\Sinei''(z) + z\,\Sinei'(z) = 0, \qquad \Sinei(0)=0, \Sinei'(0)=1
\]
the result of our algorithm shows that the Sine integral special function may be approximated with absolute error less than $10^{-100}$ on the disk $\abs{z} \leq 1$ by truncating its Taylor series at the origin to the order 74.
\end{enumerate}
\end{exmp}

\paragraph*{Outline.}

\begin{figure}
\begin{tikzpicture}[
    line/.style ={draw, very thick, ->},
    dl/.append style ={dashed},
    prop/.style={text width=26mm, text centered, font=\scriptsize},
    etiq/.style= {draw, thin, rounded corners, font=\scriptsize, inner sep=4pt, text centered},
    etiqarr/.style={-*, shorten >=-.7mm} ]
  \begin{scope}[
      start chain,node distance=9mm, every join/.style={line},
      block/.style ={rectangle, on chain, draw,
                    text width=24mm, text centered, minimum height=3em} ]
      \node[block]
        (rec1) {Recurrence};
      \node[block,join]
        (rec2) {Normalized recurrence};
      \node [block, join]
        (deq) {Differential equation};
      \node [block, join=by dl]
        (fprec) {Recurrence ${y_n {=} \sum_{j=0}^{n-1} \cdots}$};
      \node [continue chain=going below, node distance=23mm, block]
        (mfprec) {Majorant recurrence};
      \node [continue chain=going left, block, join=by dl]
        (mdeq) {``Majorant equation''};
      \node [block, join]
        (form1) {Formula for normalized rec.};
      \node[block,join]
        (form2) {Formula};
  \end{scope}

  \draw[line,dl] (fprec.320) -- (mfprec.40);

  \node [draw, below=3em of form1.south, anchor=east]
    (restes) {Bounds on tails...};
  \draw[line] ($(mdeq.west) - (5mm, 0)$) |- (restes);

  \node[etiq,above=2mm of rec1.north east, text width=3cm]
    (chvar1) {Newton polygon, \\ change of variable (§\ref{sec:asympt})};
  \draw[etiqarr] (chvar1) to[out=-50,in=100] ($(rec1.east) + (5mm, 0)$);

  \draw[line] (deq.south east) to[out=-30,in=30]
      node[right, etiq, outer sep=1mm, text width=15mm]
          {Majorant\\ series (§\ref{sec:majorants})}
      (mdeq.north east);

  \node[etiq, above=2mm of form1.40, text width=36mm]
    (col) {Coefficient extraction,\\ saddle point method (§\ref{sec:explicit})};
  \draw[etiqarr] (col) to[out=-30, in=110] ($(mdeq.west) - (5mm, 0)$);

  \node[prop,below] at(rec2.south)
    {``Poincaré type''};
  \node[prop,below] at(deq.south)
    {0 regular point, \\ finite radius of convergence};
  \node[prop,below] at(mdeq.south)
    {order 1,\\ simple solution};
  \node[prop,below] at(form1.south)
    {$\abs{\tilde u_n} \leq \alpha^n \phi(n)$};
  \node[prop,below] at(form2.south)
    {$\abs{u_n} \leq n!^{p/q} \alpha^n \phi(n)$};
\end{tikzpicture}
\caption{Outline of our bound computation method. Solid arrows represent computation steps; dashed arrows indicate proof steps without counterpart in the algorithm.}
\label{fig:outline}
\end{figure}
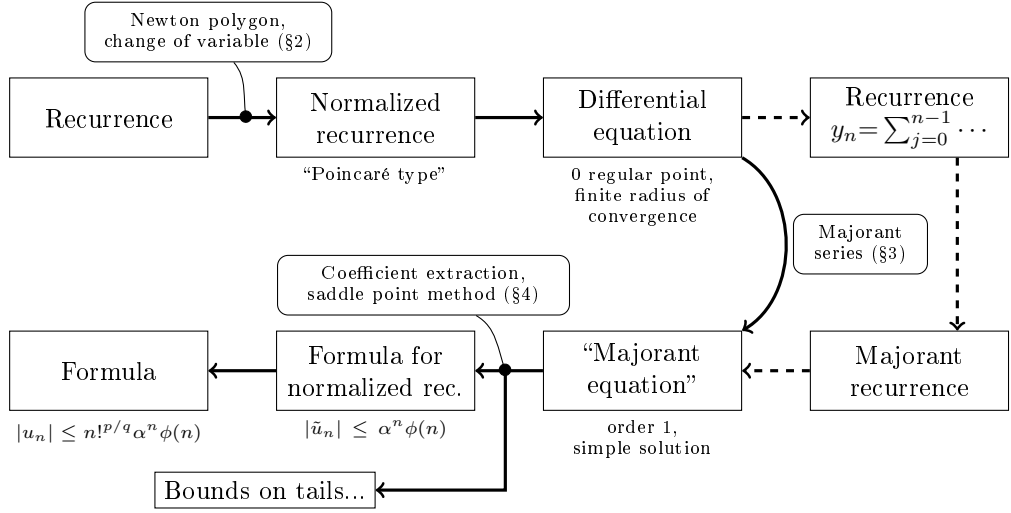

Our approach is summarized in Figure~\ref{fig:outline}.
Consider a solution $(u_n)$ of Equation~\eqref{eq:prec}. Classical methods involving Newton polygons and characteristic equations allow to extract from the recurrence relation some information on the asymptotic behaviours that $(u_n)$ may assume. We use these methods to ``factor out'' the main asymptotic behaviour, thus reducing the computation of a bound on $\abs{u_n}$ to that of a bound on a sequence of subexponential growth. This sequence is solution to a ``normalized recurrence'' computed in that step. Using the correspondence between P-recursive sequences and D-finite functions, we  encode this sequence by a differential equation satisfied by its generating function~(§\ref{sec:asympt}). Then we adapt the method of Cauchy-Kovalevskaya majorant series to bound this generating function. The key point here, in view of the requirement of asymptotic tightness, is to find a majorant whose disk of convergence extends to the nearest singularity of the equation, thus avoiding the loss of an exponential factor usually associated with the majorant series method~(§\ref{sec:majorants}). We show how to deduce several kinds of explicit bounds on $u_n$ and $\sum_n u_n z^n$ from the asymptotic behaviour and the majorant series~(§\ref{sec:explicit}). Finally, we introduce our implementation of the algorithms of this article and we briefly discuss their use in the context of high-precision numerical evaluation~(§\ref{sec:applications}).

\paragraph*{Terminology and Notations.}

We let $\Q[n]\langle S \rangle$ be the algebra of recurrence operators with polynomial coefficients, viewed as noncommutative polynomials over $\Q[n]$ in the shift operator $S : \CC^\N \to \CC^\N, (u_n)_{n\in\N} \mapsto (u_{n+1})_{n\in\N}$. Note that the sequences we consider are indexed by the nonnegative integers. Similarly, $\partial$ stands for the differentiation of formal power series, and $\Q[z]\langle\partial\rangle$ for the algebra of linear differential operators with polynomial coefficients, written with $\partial$ on the right. Noncommutative monomials are written and represented in memory with the coefficient on the left and the power of the main variable $S$ or $\partial$ on the right.

For any formal power series $u \in \CC[[z]]$, we denote by $u_n$ (or sometimes by $[z^n] u$) the coefficient of $z^n$ in $u$. Following \citet{vdH2003}, we also write
\[ u_{;n} = \sum_{k=n}^\infty u_k z^k, \quad u_{n;} = \sum_{k=0}^{n-1} u_k z^k. \]
To avoid ambiguity, most other indexed names are written using bracketed superscripts, like $p^{[0]}$ in Equation~\eqref{eq:prec}.
We use the notations of \citet{ConcreteMath} for the rising and falling factorials, namely $x^{\rising{n}}=\prod_{k=0}^{n-1} (x+k)$ and $x^{\underline{n}}=\prod_{k=0}^{n-1} (x-k)$.

In the statement of algorithms, we employ expressions such as ``set $x \geq v$'', to mean ``compute an approximation of $v$ by excess (without any precise accuracy requirement) and assign it to~$x$''.


\section{Factorial and Exponential behaviour} \label{sec:asympt} 

In this section, we collect classical results on the asymptotics of P-recursive sequences. These will both allow us to make precise statements about the tightness of the bounds we compute and serve as a guide to organise the computation in order to meet these requirements. Moreover, we state effective versions of some parts of the results, that constitute the first steps of our algorithm.

\subsection{The Perron-Kreuser theorem} \label{sec:PK}

\begin{figure}
\newcommand{\newtonpolygon}[4]{
\begin{scope}
  \node[coordinate] (topright) at (#1) {};
  \node[coordinate] (orig) at (0,0) {};
  \draw[thick, fill=gray!40] #4;
  \draw[thin, dotted] grid[step=1] (topright);
  \draw[->] (orig) -- ($(0,0)!(topright)!(1,0)$) node[right] {$S$};
  \draw[->] (orig) -- ($(0,0)!(topright)!(0,1)$) node[above] {$n$};
  \node (#2) [fit=(orig) (topright)] {};
  \foreach \coeff in {#3}
    \fill \coeff circle (2pt);
\end{scope}
}
\begin{center}
\begin{tikzpicture}[x=0.5cm,y=0.5cm]
\begin{scope}
  \newtonpolygon{3.8,1.8}{n1}%
    {(0,0), (0,1), (1,1), (2,0), (3,0)}%
    {(0,0) -- (0,1) -- (1,1) -- (3,0)}
  \node (slope label) [above=5mm of n1.40] { slope $-\kappa = -1/2$ };
  \draw [->, bend left] (slope label) to (2.5,0.25);
\end{scope}
\begin{scope}[xshift=7cm] 
  \newtonpolygon{6.8,3.8}{n2}%
    {(0,0), (0,1), (2,0), (2,1), (2,2), (2,3), (4,0), (4,1), (4,2), (4,3), (6,0), (6,1), (6,2), (6,3),}%
    { (0,0) -- (0,1) -- (2,3) -- (6,3) -- (6,0) }
\end{scope}
  \draw[-angle 90, thick]
    ($(n1.south east) + (0.5cm,0.3cm)$) to
      node[auto, text width=3cm, text centered] {
        $\tilde u_n = \psi_n u_n$ \\
        $(n+q)^p \psi_{n+q}=\psi_n$
      }
    ($(n2.south west) + (-0.5cm,0.3cm)$);
  \node[below=0cm of n1, text width=2.7cm]
    {$\scriptstyle{u_{n+3} + u_{n+2} + n u_{n+1}}$ ${}\qquad \scriptstyle{+ (n+1) u_n =0}$};
  \node[below=0cm of n2, text width=4.7cm] {
    $\scriptstyle{(n+6)(n+4)\tilde u_{n+6}+2(n+4)(n+1)\tilde u_{n+4}}$ ${}\qquad \scriptstyle{-(n^2-n-5)\tilde u_{n+2}-(n+1)\tilde u_n=0}$};
\end{tikzpicture}
\caption{Newton polygons of recurrence operators, before and after normalization.}
\label{fig:newtonpolygons}
\end{center}
\end{figure}

A linear recurrence relation
\begin{equation} \label{eq:PKrec}
p^{[s]}(n)  u_{n+s} + \dots + p^{[1]}(n)  u_{n+1}  + p^{[0]}(n)  u_{n} = 0,
\end{equation}
or the corresponding operator $\sum p^{[k]} S^k$, is called \emph{nonsingular} when $p^{[s]}(n) \neq 0$ for all $n \in \N$. It is called \emph{reversible}
when~$p^{[0]}(n)\neq0$ for all $n \in \N$. 

Assume that the coefficients $p^{[k]}(n)$, $k=0,\dots,s$ of \eqref{eq:PKrec} are sequences such that $p^{[k]}(n)\sim_{n \to \infty} c_k n^{d_k}$ for some $c_k \in \CC, d_k \in \Z$ (for instance, they are rational functions of $n$). If $(u_n)$ is a solution of~\eqref{eq:PKrec} with $u_{n+1}/u_n \sim_{n \to \infty} \lambda n^\kappa$ then for  the recurrence equation to hold asymptotically, the maximum value of $d_k + k \kappa$ for $k = 0, \dots, s$ must be reached at least twice, so that the corresponding terms can cancel. This means that $-\kappa$ must be among the slopes of the edges of the \emph{Newton polygon} of the equation.

The Newton polygon of~\eqref{eq:PKrec} is the upper convex hull of the points $(k, d_k) \in \R^2$, $k=0, \dots, s$ (see Figure~\ref{fig:newtonpolygons}). If~$e$ is an edge of the polygon,  we denote by $-\kappa(e)$ its slope. If $(t, d_t)$ is the leftmost point of~$e$, then the algebraic equation
\begin{equation} \label{eq:char}
\chi_e(\lambda) = \sum_{(k,d_k) \in e} c_k \lambda^{k-t} = 0
\end{equation}
is called the \emph{characteristic equation} of $e$. Observe that the degrees of the characteristic equations sum up to the order $s$ of the recurrence.

\begin{thm}[Poincaré, Perron, Kreuser] \label{thm:PK}
For each edge~$e$ of the Newton polygon of~\eqref{eq:PKrec}, let $\lambda_{e,1}, \lambda_{e,2}, \dots$ be the solutions of the characteristic equation $\chi_e$, counted with multiplicities.
\begin{enumerate}[(a)]
\item \label{item:Poincare} If for each $e$, the moduli $\abs{\lambda_{e,1}}, \abs{\lambda_{e,2}}, \dots$ are pairwise distinct, then any solution $(u_n)$ that is not ultimately~0 satisfies $u_{n+1}/u_n \sim_{n \to \infty} \lambda_{e,i} n^{\kappa(e)}$ for some $e$ and $i$.
\item \label{item:PerronKreuser1} If moreover \eqref{eq:PKrec} is reversible, then it admits a basis of solutions $(u^{[e,i]})_{e, 1 \leq i \leq \deg \chi_e}$ such that 
\begin{equation} \label{eq:PKbasis1}
\frac{u^{[e,i]}_{n+1}}{u^{[e,i]}_n} \sim_{n \to \infty} \lambda_{e,i} n^{\kappa(e)}.
\end{equation}
\item \label{item:PerronKreuser2} If there exists~$e$ and~$i\neq j$ such that $\abs{\lambda_{e,i}} = \abs{\lambda_{e,j}}$, results analogous to (\ref{item:Poincare}) and (\ref{item:PerronKreuser1}) hold with the weaker conclusion
\begin{equation} \label{eq:PKbasis2}
\limsup_{n\to\infty} \abs{\frac{u_n^{[e,i]}}{n!^{\kappa(e)}}}^{1/n} \!\!= \abs{\lambda_{e,i}}.
\end{equation}
\end{enumerate}
\end{thm}

\begin{defn}[Normalized Recurrences] \label{def:normalized}
If all the edges have nonnegative slope (i.e., if after dividing~\eqref{eq:PKrec} by $p^{[s]}$, each coefficient tends to a finite limit as $n \to \infty$), the recurrence is said to be of \emph{Poincaré type}. In that case, we call it (and the corresponding operator) \emph{normalized} if the Newton polygon has a horizontal edge.
\end{defn}

Thus a normalized recurrence is one whose ``fastest growing'' solution has purely exponential (as opposed to factorial) growth.

Item (\ref{item:Poincare}) above is known as Poincaré's theorem~\citep{Poinca1885}; Items (\ref{item:PerronKreuser1}) and (\ref{item:PerronKreuser2}) are Perron's theorem~\citep{Perron1909a, Perron1909, Perron1921} in the case of recurrence relations of Poincaré type, and the Perron-Kreuser theorem~\citep{Perron1910, Kreuse1914} in the general case. In addition to the original works, we refer to \citet{Meschk1959} and \citet{Guelfo1963} for accessible proofs of Poincaré's and Perron's theorems. Various further extensions and refinements of these results are available, see, e.g., \citet{Schaefk1965}, \citet{KooTij1990}, \citet{Pituk1997}, \citet{BusBus2005}, and the references therein.

In other words, the Perron-Kreuser theorem states that~\eqref{eq:PKrec} admits a basis of solutions of the form given by Theorem~\ref{thm:PK} in some neighborhood of infinity. The assumption that~\eqref{eq:PKrec} is reversible ensures that any solution near infinity extends to a solution defined on the whole set of nonnegative integers. 

\subsection{Dominant Singularities} \label{sec:domsing}

If $P$ is a polynomial, we denote by $\ord(\zeta,P)$ the multiplicity of $\zeta$ as a root of $P$. We call \emph{dominant roots} of $P$ those of highest multiplicity among its \emph{nonzero} roots of \emph{smallest} modulus. We denote by $\delta(P)$ and $\ord_\delta(P)$ their modulus and multiplicity, respectively. By convention, the dominant root of a monomial is $\infty$. We call \emph{dominant poles} of a rational function the dominant roots of its denominator; and \emph{dominant singularities} of a differential operator with polynomial coefficients the dominant roots of its leading coefficient.

Besides standard symbolic manipulation routines, we  assume that we have at our disposal a few operations on real algebraic numbers represented using the notation $\delta(P)$, namely a function that decides, given $P, Q \in \Q[z]$, whether $\delta(P) < \delta(Q)$, $\delta(P) = \delta(Q)$ or $\delta(P) > \delta(Q)$ and a procedure to compute arbitrarily good lower approximations of $\delta(P)$. The comparison can be based on a symbolic-numeric approach as in \citep{GouSal1996}.  Modern polynomial root finders such as MPSolve~\citep{BiniFiorentino2000} or those of major computer algebra systems provide the required numerical evaluation features---and much more. Since we are interested only in $\delta(P)$ as opposed to all roots of $P$, we may also use a simple procedure based on Graeffe's method \citep[see, e.g.,][§14]{Sch1982} if no general polynomial solver is available. More generally, most steps of Algorithms~\ref{algo:boundratpoly} and~\ref{algo:boundnormaldiffeq} involving no precise accuracy requirement may be implemented using interval arithmetic or floating-point arithmetic with careful rounding instead of symbolically.

\begin{rem}
Although we work over $\Q$ all along this paper for clarity, we expect that most results adapt without difficulty to any ``sufficiently effective'' subfield of $\CC$. However, the way to perform the basic operations we assume available in this section (as well as the details of some algorithms, especially Algorithm~\ref{algo:boundratpoly} below) may differ.
\end{rem}

\subsection{Generic Growth of the Solutions} \label{sec:growth}

\begin{algorithm2e}[t]
\caption{Factorial and exponential behaviour} \label{algo:growth}
\Function{\Asympt{$\sum_{k=0}^s b^{[k]}(n) S^k \in \Q[n]\langle S\rangle$}}{%
    $\kappa \gets \max_{k=0}^{s-1} \frac{\deg b^{[k]} - \deg b^{[s]}}{s-k}$
    \;
    $P_\alpha \gets \sum_{\ell=0}^s b^{[s-\ell]}_{d+\ell\kappa} z^\ell$ where $d = \deg b^{[s]}$
    \;
    return $(\kappa, P_\alpha)$%
}
\end{algorithm2e}

Let $R \in \Q[n]\langle S \rangle$ be a nonsingular reversible operator of order $s$. Then any solution of the recurrence relation $R·u=0$ is uniquely determined by its initial values $(u_0, \dots, u_{s-1}) \in \CC^s$. Accordingly, we say that an assertion is true \emph{for a generic solution} of $R·u=0$, or \emph{for generic initial values}, if it is true for any solution $u$ such that $(u_0, \dots, u_{s-1}) \in \CC^s \setminus V$ where $V$ is a \emph{proper} linear subspace of $\CC^s$.

Theorem~\ref{thm:PK} implies that the factorial and exponential asymptotic behaviour of the ``fastest growing'' solutions is determined by the dominant singularities of $R$. We use Algorithm~\ref{algo:growth} to extract this asymptotic behaviour, which is in fact that of a generic solution of $R·u=0$, as stated by Proposition~\ref{prop:tightness} below.

\begin{prop}[Factorial and Exponential Growth] \label{prop:tightness}
Write $R$ as $\sum_{k=0}^s b^{[k]}(n) S^k \in \Q[n]\langle S\rangle$ and assume $b^{[k]}b^{[s]} \neq 0$ for some~$k\in\{0,\dots,s-1\}$. Algorithm~\ref{algo:growth} computes $(\kappa, P_\alpha) = \Asympt{R}$ such that for any solution $(u_n)$ of $R \cdot u = 0$,
\begin{equation} \label{eq:limsup}
\limsup_{n \to \infty} \abs{\frac{u_n}{n! ^\kappa}}^{1/n} \leq \alpha \qquad \text{where } \alpha = \frac 1 {\delta(P_\alpha)},
\end{equation}
with equality in the generic case.
\end{prop}

\begin{pf}
The inequality follows from Theorem~\ref{thm:PK} since $-\kappa$ is the slope of the rightmost edge $e$ of the Newton polygon of $R$ and $P_\alpha$ is the reciprocal polynomial of $\chi_e$. It remains to show that equality holds for generic initial values. Let $V = \ker R \subset \CC^\N$. Also by Theorem~\ref{thm:PK}, there exists $u^{[1]} \in V$ such that
\[
\limsup_{n \to \infty} \abs{\frac{u^{[1]}_n}{n! ^\kappa}}^{1/n} = \alpha.
\]
This can be extended to a basis $u^{[1]}, \dots, u^{[s]}$ of $V$. Let $u = \sum_k \lambda^{[k]} u^{[k]} \in V$. By construction of $\kappa$ and $\alpha$, we have the inequality $\limsup \abs{u_n/n!^\kappa}^{1/n} \leq \alpha$. Up to extraction of a subsequence we can assume (i) that $u^{[1]}_n$ does not vanish, (ii) that $\mathopen|u^{[1]}_n/n!^\kappa\mathclose|^{1/n} \to \alpha$ and (iii) that there exists $\beta \leq \alpha$ such that $\abs{u_n/n!^\kappa}^{1/n} \to \beta$ as $n \to \infty$. Then 
\[ \abs{\lambda^{[1]} + \lambda^{[2]} \frac{u^{[2]}_n}{u^{[1]}_n} + \dots + \lambda^{[s]} \frac{u^{[s]}_n}{u^{[1]}_n}}^{1/n} \to \frac\beta\alpha, \]
so that $\beta = \alpha$ unless
\[ \frac{\lambda^{[2]} u^{[2]}_n + \dots + \lambda^{[s]} u^{[s]}_n}{u^{[1]}_n} \to - \lambda^{[1]}, \]
which does not happen for generic $\lambda^{[k]}$.
\end{pf}

Accordingly tighter results hold if the assumptions of Theorem~\ref{thm:PK}(\ref{item:PerronKreuser1}) are fulfilled.

\subsection{Generating Function and Associated Differential Equation} \label{sec:rectodiffeq}

\begin{algorithm2e}[t]
\caption{Recurrence to normalized differential equation} \label{algo:normalize}
\Function{\RecToDiffeq{$R = \sum_{k=0}^s b^{[k]} S^k \in \Q[n] \langle S \rangle$}}{
$g \gets \Pi / \gcd(b^{[s]},\Pi)$ where $\Pi = \prod_{k=1}^s (n+k)$
\;
$\sum_{k=0}^s c_{kj} n^j S^k \gets g \, R$
\Rem*[f]{thus $R = \sum_{k=0}^s c_{kj} S^k (n-k)^j$}
\;
expand $\sum_{k=0}^s \sum_j c_{kj} z^{s-k} (\theta-k)^j$ as $D = \sum_{k=0}^r a^{[k]} \theta^k$
\;
return $D$
}
\Function{\NormalizedDiffeq{$R \in \Q[n] \langle S \rangle, \kappa \in \Q$}}{
$p/q \gets \kappa$ (in irreducible form, with $(p,q)=(0,1)$ if $\kappa=0$)
\;
compute the symmetric product $\hat R = \sum_{k=0}^{qs} \hat b^{[k]}(n) S^k$ of $R$ and $(n+q)^p S^q - 1$ \;  
\Rem{see, e.g., \citet[§6.4]{Stanley1999}}
\;
return \RecToDiffeq{$\hat R$}
}
\end{algorithm2e}

Consider again a nonsingular recurrence operator $R = \sum_{k=0}^s b^{[k]} S^{k} \in \Q[n]\langle S \rangle$ (with $b^{[0]}, b^{[s]} \neq 0$). Using the \emph{Euler derivative} $\theta = z \frac{d}{dz}$, it is classical that the generating series $u(z)$ of $u \in \ker R$ cancels the associated differential operator $D = \sum_{k=0}^r a^{[k]} \theta^k \in \Q[z] \langle \theta \rangle$ computed by $\RecToDiffeq$ (Algorithm~\ref{algo:normalize})%
\footnote{Actually, the classical translation of recurrence operators to differential operators uses $g=1$. The multiplication by $g$ in our version comes from our choice to use sequences indexed by $\N$ rather than $\Z$.}%
. Dividing out by $a^{[r]}$, this rewrites
\begin{equation} \label{eq:theta}
\Bigl( \theta^r + \frac{a^{[r-1]}}{a^{[r]}}\, \theta^{r-1} + \dots + \frac{a^{[1]}}{a^{[r]}} \theta + \frac{a^{[0]}}{a^{[r]}} \Bigr) \cdot u = 0.
\end{equation}
A point $z_0 \in \CC$ is a \emph{regular point} of~\eqref{eq:theta} if any solution $u$ has polynomial growth $u(z) = 1/\abs{z-z_0}^{O(1)}$ as $z \to z_0$ in a sector with vertex at $z_0$. Regular points encompass \emph{ordinary points}, where the equation is nonsingular and thus has analytic solutions by Cauchy's theorem, and \emph{regular singular points}. Fuchs' criterion \citep[see, e.g.,][§15.3]{Ince1956} states that~0 is a regular point if and only if for all $k$, the coefficient $a^{[k]}/a^{[r]}$ of~\eqref{eq:theta} is analytic at $0$, while $z_0\neq0$ is a regular point if and only if each $a^{[k]}/a^{[r]}$ has a pole of order at most $r-k$ in $z_0$. (This criterion still holds if the $a^{[k]}/a^{[r]}$ are replaced by meromorphic functions.)

\begin{lem} \label{lem:normalization}
If $R$ is normalized (Definition~\ref{def:normalized}), then the origin is a regular point of $D$, and the reciprocal polynomial of the leading term $a^{[r]}$ of $D$ is the characteristic equation of the horizontal edge of the Newton polygon of $R$. 
\end{lem}

\begin{pf}
Using the notations of the function \RecToDiffeq{} in Algorithm~\ref{algo:normalize}, let $d^{[k]} = \deg b^{[k]}$ for all $k$, and $m = \deg g$. Thus $r = \max_{k=0}^s d^{[k]} + m$. The leading term of $\theta^j z^{-k}$ as an operator in $\theta$ with Laurent polynomial coefficients is $z^{-k} \theta^j$, hence $a^{[r]}(z) = \sum_{k=0}^s c_{kr} z^{s-k}$. The condition that $R$ is normalized translates into $d^{[s]} = \max_{k=0}^{s-1} d^{[k]}$, that is, $d^{[k]}=d^{[s]}=r-m$ for some $k < s$. It follows that $a^{[r]}(0) = c_{s r} \neq 0$, hence $0$ is a regular point by Fuchs' criterion. Finally, if $R$ is normalized and if $e$ is the edge of its Newton polygon such that $\kappa(e)=0$, then the general expression
\[ \chi_e(\lambda) = \lambda^{-t} \sum_{\substack{d^{[k]}+k\kappa(e) \\=d^{[s]}+s\kappa(e)}} a_{k,d^{[k]}} \lambda^k \]
(where $t$ is such that $\chi_e(0) \neq 0$) simplifies to $\chi_e(\lambda)=\lambda^{-t} \sum_{d^{[k]}=r} a_{k,r} \lambda^k$.
\end{pf}

In the general case, we normalize $R$ by a change of unknown sequence preserving P-recursiveness before we compute the associated differential equation. This is described in the next proposition. Figure~\ref{fig:newtonpolygons} gives an example of normalization of recurrence operators and of its action on their Newton polygons.

\begin{prop} \label{prop:diffeqtorec}
Let $R \in \Q[n]\langle S \rangle$ be nonsingular, reversible, with nonzero constant coefficient with respect to $S$. Let $(p/q, P_\alpha) = \Asympt(R)$ as computed by Algorithm~\ref{algo:growth}, and assume that $\delta(P_\alpha) < \infty$. Algorithm~\ref{algo:normalize} computes a normalized differential operator $D = \NormalizedDiffeq(R, p/q)$ that cancels $\tilde u(z) = \sum_{n=0}^\infty \psi_n u_n z^n$ for all sequences $\psi$ and $u$ such that
\[ (n+q)^p \psi_{n+q}=\psi_n \quad \text{and} \quad R \cdot u = 0. \]
The origin is a regular point of $D$, and the modulus of the dominant singularities of $D$ equals $\delta(P_\alpha)$.
\end{prop}

\begin{pf}
Let $\alpha = 1/\delta(P_\alpha)$.
Let $(u^{[1]}, \dots, u^{[s]})$ be a basis of $\ker R$ having the asymptotic behaviours given by~\eqref{eq:PKbasis2}. In particular $\limsup_{n \to \infty} \mathopen| u^{[k]}/n!^{p/q} \mathclose|^{1/n} \leq \alpha$ for all $k$. Let $(\psi^{[0]}, \dots, \psi^{[q-1]})$ be the basis of solutions to $(n+q)^p \psi_{n+q} = \psi_n$ corresponding to the initial values $\psi^{[i]}_j = \delta_{ij}$ for $0 \leq i, j < q$,
where $\delta_{ij}$ is the Kronecker symbol. Algorithm~\ref{algo:normalize} constructs $\hat R$ such that for $N$ large enough, the $sq$ sequences $(\psi^{[j]}_n u^{[k]}_n)_{n \geq N}$ generate $\{ \hat u \; | \; (\hat R \cdot \hat u)_n = 0 \text{ for } n \geq N \}$. For all $j$ and $k$, $\limsup \mathopen| \psi^{[j]}_n u^{[k]}_n \mathclose|^{1/n} \leq \alpha$. Assume that $\hat u = \sum_{j,k} \lambda^{[j,k]} \psi^{[j]} u^{[k]}$ is solution to $\hat R \cdot \hat u = 0$ in some neighborhood of infinity. Then $\limsup \mathopen| u_n \mathclose|^{1/n} \leq \alpha$ (indeed, if $\epsilon > 0$, then $\abs{u_n} \leq (\alpha+\epsilon)^n$ for $n$ large enough). On the other hand $\limsup \mathopen| \psi^{[j]}_n u^{[k]}_n \mathclose|^{1/n} = \alpha$ for at least one $(j,k)$. Hence, by Theorem~\ref{thm:PK}, the operator $\hat R$ is normalized and the largest modulus of a root of the characteristic equation associated to the horizontal edge of its Newton polygon is $\alpha$. Applying Lemma~\ref{lem:normalization} concludes the proof.
\end{pf}

In the sequel, we will choose as normalizing sequence the solution to $(n+q)^p \psi_{n+q}=\psi_n$ given by
\[ \psi_n = q^{-\frac p q n} \Gamma(n/q+1)^{-p}. \]
Observe that $(\psi_n)_{n \in \N}$ is monotone: indeed, the function $x \mapsto q^x \Gamma(x+1)$ is increasing for $x \geq 0$ as soon as $\log q > \gamma$ (the Euler–Mascheroni constant), and the remaining case $q=1$ is obvious.


\section{Subexponential Behaviour: Majorant Series Computation} \label{sec:majorants}

The results of the previous section allow us to compute the generic factorial and exponential asymptotic behaviour of solutions of a linear recurrence relation with polynomial coefficients. We now turn to the computation of a bound for the remaining subexponential factor of a particular solution.

\subsection{Majorant Series and the Cauchy-Kovalevskaya Method} \label{sec:CK}

The main tool we use is a variant of the Cauchy-Kovalevskaya majorant series method, which usually serves to establish the convergence of formal series solutions to differential and partial differential equations, but may also be applied to obtain explicit bounds on the tails of these solutions (see also~\S\ref{sec:num} for more on this).

\begin{defn}[Majorant series] \label{def:majorant}
A formal power series $v \in \R_+[[z]]$ is a \emph{majorant series} of $u \in \CC[[z]]$, and we write $u \maj v$, if $\abs{u_n} \leq v_n$ for all $n \in \N$.
\end{defn}

In particular, the disk of convergence of $v$ is contained in that of $u$, and if $z$ lies inside the disk of convergence of $v$, we have that $\abs{u_{n;}(z)} \leq v_{n;}(\abs z)$  for all $n \geq 0$. Other immediate properties of majorant series are summarized in the following lemma.

\begin{lem} \label{lem:majorants}
Assume that $u, u^{[1]}, u^{[2]} \in \CC[[z]]$, $v, v^{[1]}, v^{[2]} \in \R_+[[z]]$ are such that
$u \maj v$, $u^{[1]}\maj v^{[1]}$ and $u^{[2]} \maj v^{[2]}$. 
Then
\[
\frac{du}{dz} \maj \frac{dv}{dz}; \quad
u^{[1]}+u^{[2]} \maj v^{[1]}+v^{[2]}; \quad
u^{[1]}u^{[2]} \maj v^{[1]}v^{[2]}; \quad
u^{[2]} \circ u^{[1]} \maj v^{[2]} \circ v^{[1]}
\]
where in the last inequality it is assumed that $u^{[1]}(0)=v^{[1]}(0)=0$.
\end{lem}

In the neighborhood of an ordinary point, majorant series for the coefficients of a differential equation like~\eqref{eq:dfinite} give rise to similar majorants for the solutions. Indeed, if 
\[
\left\{
\begin{aligned}
u^{(r)} &= a^{[r-1]} u^{(r-1)}  + \cdots + a^{[0]} u  \\
v^{(r)} &= b^{[r-1]} v^{(r-1)}  + \cdots + b^{[0]} v
\end{aligned}
\right.
\qquad
\mathopen|u(0)\mathclose| \leq v(0), \dots, \mathopen|u^{(r-1)}(0)\mathclose| \leq v^{(r-1)}(0)
\]
where $a^{[k]}, b^{[k]}$ are analytic functions at $0$ such that $a^{[k]} \maj b^{[k]}$ for all $k$, then by induction $u \maj v$. This result does not hold if one of the $a^{[k]}$ has a pole at $0$; however, the method may be adapted to the case where $0$ is a regular singular point of the differential equation. We give one way to do this in §\ref{sec:dfinite}; for a more complete introduction to the ``usual'' Cauchy-Kovalevskaya method in the ODE setting covering the regular singular case, see \citet{MezPin1998}, and for a more general statement along these lines, see \citet[Proposition~3.7]{vdH2003}. In any case, the first step for obtaining majorant series for the solutions of a differential equation using the Cauchy-Kovalevskaya method is to compute majorants for its coefficients, which in the case we are interested in are rational functions.

\subsection{Bounds for Rational Functions} \label{sec:rat}

\begin{algorithm2e}[t]
\caption{Tight majorant series for rational functions} \label{algo:boundratpoly}
\Function{$\BoundRatpoly(r=N/D \in \Q(z), P_\alpha \in \Q[z], m \in \N^\ast)$}{
let $A + B/D = r$ with $A, B \in \Q[z]$.
\;
compute the squarefree factorization $D = D_1 D_2^2 \cdots D_k^k$ of $D$
\;
compute the coefficients $h_{i,d} \in \Q[\zeta]$ of the partial fraction decomposition
$ \frac{B(z)}{D(z)} = \sum_{i=1}^k \sum_{D_i(\zeta)=0} \sum_{d=1}^{i} \frac{h_{i,d}(\zeta)}{(\zeta-z)^d} $
\Rem*[f]{See, e.g., \citet[§2.7]{Bronstein2005}}
\;
\For{$i = 1, \dots, k$}{
    \For{$d = 1, \dots, i$}{
        set $c_{i,d} \geq  \sum_{D_i(\zeta)=0} \abs{ h_{i,d}(\zeta) \zeta^{-d} }$
        \nllabel{algo:boundratpoly:sym}
}
}
set 
$N_0 \geq \max \bigl(1, 1+\deg A, \max_{i=m+1}^k \frac{i-m}{\log(\delta(D_i)/\delta(P_\alpha))}\bigr)$
\;
let $t(n) = \sum_{i=1}^k \sum_{d=0}^{i-1} c_{i,d} \frac{(n+1)^{\rising{d-1}}}{(n+1)^{\rising{m-1}}} \left( \delta(P_\alpha)/\delta(D_i) \right)^{n}$
\;
compute the truncated series $r_{;N_0}(z) = \sum_{n=0}^{N_0-1} r_n z^n$ \nllabel{algo:boundratpoly:ploum}
\;
set
$h(N_0) \geq \max_{n=0}^{N_0-1} \bigl( \abs{r_n} / \bigl({\binom{n+m-1}{m-1} \delta(P_\alpha)^n} \bigr) \bigr)$
\nllabel{algo:boundratpoly:plim}
\;
return an approximation by excess of $\max\big(h(N_0), t(N_0)\big)$
}
\end{algorithm2e}

Consider a rational function $r(z) = N(z)/D(z) = \sum r_n z^n$, $D(0) \neq 0$. The sequence $(r_n)$ satisfies a linear recurrence relation with constant coefficients, whose characteristic polynomial is the reciprocal polynomial of $D$. This recurrence can be solved by partial fraction decomposition of $r$, yielding the explicit expression (recall that $x^{\underline n}$ and $x^{\overline n}$ denote respectively the falling and rising factorials)
\begin{equation} \label{eq:recsol}
r_n = \sum_{D(\zeta)=0} \sum_{d=1}^{\ord(\zeta,D)} h^{[\zeta,d]} · (n+1)^{\rising{d-1}} · \zeta^{-n},\qquad n \ge \max(0,\deg N-\deg D+1),
\end{equation}
with $h^{[\zeta,d]} \in \Q(\zeta)$.
We are now aiming at a bound of the form~$|r_n|\le M \delta(D)^{-n}n^{\ord_\delta D}.$ In view of later needs, Algorithm~\ref{algo:boundratpoly} takes as input a polynomial~$P_\alpha$ and a positive integer~$m$. It returns a bound of the form~$r(z) \maj M(1-\alpha z)^{-m}$, where $\alpha=1/\delta(P_\alpha)$. In particular, when~$P_\alpha=D$ and~$m=\ord_\delta(D)$ this bound is tight.

To compute a suitable $M$, we start with the right-hand side of~\eqref{eq:recsol} divided by 
\[ b_n = [z^n]\frac{1}{(1-\alpha z)^m} = (n+1)^{\rising{m-1}} · \alpha^n.\]
By applying the triangle inequality, we get a sum $t(n)$ of terms of the form
\[ c \; \frac{(n+1)^{\rising{d-1}}}{(n+1)^{\rising{m-1}}} \; \lambda^n \]
where $0\le c$, $0 < \lambda \leq 1$, and $m<d$ only if $\lambda < 1$. Such a term is decreasing for $n \geq 1$ if $d \leq m$ and for $n \geq (d-m)/\log(1/\lambda)$ otherwise. We compute an index $N_0$ starting from which the inequality $\abs{r_n/b_n} \leq t(n)$ is guaranteed to hold and $t(n)$ is guaranteed to be decreasing; then we adjust $M$ from the explicit values of the first~$N_0$ coefficients and bounds on the tails.

For this last part, consider the squarefree decomposition $D = D_1 D_2^2 \cdots D_k^k$. If $\zeta$ is a root of $D_i$, then each $h^{[\zeta,d]}$ may in fact be written $h^{[\zeta,d]} = h_{i,d}(\zeta)·\zeta^{-d}$ for some polynomial $h_{i,d} \in \Q[\zeta]$ depending only on $D_i$ and $d$. Moreover, in this expression, $\abs{\zeta}^{-1}$ may be bounded by $\delta(D_i)^{-1}$. Hence we have
\begin{equation}
\begin{aligned} 
\abs{\frac{r_n}{b_n}}
  &= \left|
    \alpha^{-n} 
    \sum_{i=1}^k \sum_{D_i(\zeta)=0} \sum_{d=0}^{i-1}
       h_{i,d}(\zeta) \zeta^{-d}
       \frac{(n+1)^{\rising{d-1}}}{(n+1)^{\rising{m-1}}}
       \zeta^{-n} \right| \\
  &\leq \sum_{i=1}^k 
    \sum_{d=0}^{i-1} \; 
    \Biggl( \sum_{D_i(\zeta)=0} \Bigl| \frac{h_{i,d}(\zeta)}{\zeta^{d}}  \Bigr| \Biggr) \;
    \frac{(n+1)^{\rising{d-1}}}{(n+1)^{\rising{m-1}}} \;
    \bigl(\alpha \, \delta(D_i) \bigr)^{-n}. 
\label{eq:triangle}
\end{aligned}
\end{equation}
We may take for $t(n)$ the right-hand side of~\eqref{eq:triangle}, or even a suitable numerical approximation.
To deal with the sum in parentheses, we may bound $\zeta^{-d} h_{i,d}(\zeta)$ term-by-term, replacing once again $\zeta^\ell$ by $\delta(D_i)^\ell$ or $\delta(\zeta^{\deg D_i} P_i(1/\zeta))^{-\ell}$ depending on the sign of $\ell$. We may also simply compute low-precision enclosures of the roots of $D_i$ and then use interval arithmetic.

The complete procedure is summarized in Algorithm~\ref{algo:boundratpoly}. We have thus proved the following.

\begin{prop}
Given $r = N/D \in \Q(z)$ (in irreducible form), $P_\alpha \in \Q[z]$, and $m \in \N^\ast$, such that $0 < \delta(P_\alpha) \leq \delta(D)$ and $\delta(P_\alpha)=\delta(D)$ only if $m \geq \ord_\delta D$, Algorithm~\ref{algo:boundratpoly} computes $M = \BoundRatpoly(r, P_\alpha, m) \in \Q_+$ satisfying $r(z) \maj M (1-z/\delta(P_\alpha))^{-m}$.
\end{prop}

To improve $M$, we may loop over lines~\ref{algo:boundratpoly:ploum} and~\ref{algo:boundratpoly:plim} of Algorithm~\ref{algo:boundratpoly}, doubling $N_0$ each time, until $N_0$ or $t(N_0) - h(N_0)$ reaches some specified value. 

\subsection{Bounds for D-finite Functions} \label{sec:dfinite}

We now apply the Cauchy-Kovalevskaya method to deduce a majorant series for~$u(z)$ from the asymptotic behaviour of $(u_n)$ obtained in~§1 and majorant series for the coefficients of an associated differential equation. The majorant series we obtain is ``simpler'' than~$u(z)$ in the sense that it always satisfies a differential equation of order~1.

By Fuchs' criterion, we may isolate the constant term of each coefficient of~\eqref{eq:theta}, giving
\begin{equation} \label{eq:rwdeq}
Q(\theta) \cdot u = z (\tilde a^{[r-1]} \theta^{r-1} + \cdots + \tilde a^{[1]} \theta + \tilde a^{[0]}) \cdot u,
\end{equation}
where $Q \in \Q[X]$ is a monic polynomial of degree $r$ and the $\tilde a^{[k]}$ are rational functions of~$z$. Let $m_k \in \N$ be the maximum multiplicity of a point of the circle $\abs z = \delta(P_\alpha)$ as a pole of $\tilde a^{[k]}$ and let $T = \max(0, \max_{k=0}^{r-1} (m_k-r+k))$. We emphasize that, although Algorithm~\ref{algo:boundnormaldiffeq} takes $P_\alpha$ as input, the whole point of the method is that $\delta(P_\alpha)$ may indeed equal the modulus of the dominant singularities of $D$. In that case, the integer $T$ is sometimes called the Malgrange irregularity of these singularities~\citep[see][]{Malgrange1974}, and by Fuchs' criterion again, $T=0$ if and only if the dominant singularities are all regular. Using Algorithm \ref{algo:boundratpoly}, we compute bounds of the form
\begin{equation} \label{eq:ratpolyresults}
\tilde a^{[k]} \maj \frac{M^{[k]}}{(1-\alpha z)^{r-k+T}}
\qquad 
\text{i.e.,} \quad
\abs{\tilde a^{[k]}_n} \leq  M^{[k]} \, \binom{n+r-k+T-1}{r-k+T-1} \alpha^n
\end{equation}
for the coefficients of the equation, with~$\alpha=1/\delta(P_\alpha)$ as usual (lines~\ref{algo:boundnormaldiffeq:majcoeff:start}--\ref{algo:boundnormaldiffeq:majcoeff:end} of Algorithm~\ref{algo:boundnormaldiffeq}).

\begin{algorithm2e}[t]
\caption{Majorant series for normalized D-finite functions} \label{algo:boundnormaldiffeq}
\Function{\BoundNormalDiffeq{$\sum_{k=0}^r a^{[k]} \theta^k \in \Q[z]\langle\theta\rangle, P_\alpha \in \Q[z], u_{;\cdot}$}}{
\For{$k=0,\dots,r-1$}{
    $c^{[k]} \gets (a^{[k]}/a^{[r]})_{z=0}$ (or fail with error ``$0$ should be a regular point'')
    \;
    $\tilde a^{[k]} \gets (a^{[k]}/a^{[r]} - c^{[k]})/z$
}
$T \gets \max\{ 0; \ord_\delta(\den \tilde a^{[k]})-r+k \mathrel{|} 0 \leq k < r-1 \text{ and } \delta(\den \tilde a^{[k]}) = \delta(P_\alpha) \}$.
\;
\For{$k=0,\dots,r-1$\nllabel{algo:boundnormaldiffeq:majcoeff:start}}{
    $M^{[k]} \gets \BoundRatpoly(\tilde a^{[k]}, P_\alpha, T+r-k)$
    \Rem{thus $\tilde a^{[k]} \maj M^{[k]} (1-\alpha z)^{-T-r+k}$}
    \nllabel{algo:boundnormaldiffeq:majcoeff:end}
}
$M \gets \max_{k=0}^{r-1} M^{[k]} / \binom{r-1}{k}$
\;
compute $K \in \N^\ast$ such that $K \geq 2 M \delta(P_\alpha)$
\;
starting with $N_2=1$, double $N_2$ until $\sum_{k=0}^{r-1} \abs{c^{[k]}} N_2^k < (1- M \delta(P_\alpha)/K) \, N_2^r$
\;
compute $u_{;N_2+1}$ and $v_{;N_2+1}$ where $v$ is given by~\eqref{eq:normalv} with $A=1$
\nllabel{algo:boundnormaldiffeq:A:start}
\;
$A \gets \max_{n=0}^{N_2} \abs{u_n}/v_n$
\nllabel{algo:boundnormaldiffeq:A:end}
\;
return $(T, K, A)$
}
\end{algorithm2e}

Extracting the coefficient of $z^n$ in \eqref{eq:rwdeq}, we get
\begin{equation}
Q(n)\, u_n = \sum_{j=0}^{n-1} \sum_{k=0}^{r-1} \tilde a^{[k]}_{n-1-j} j^k u_j.
\end{equation}
Since $Q$ is monic, let $N_1$ be such that $Q(n) > 0$ for $n \geq N_1$; then by~\eqref{eq:ratpolyresults}, for such $n$,
\begin{equation} \label{eq:absrec}
Q(n)\, \abs{u_n} 
\leq \sum_{j=0}^{n-1} \sum_{k=0}^{r-1} M^{[k]}  \binom{n{-}1{-}j+r{-}k{+}T-1}{r{-}k{+}T-1} \alpha^{n-1-j} j^k \abs{u_j}.
\end{equation}

\begin{lem}[Reduction from order $r$ to order $1$] \label{lem:M}
Let $M = \max_{k=0}^{r-1} M^{[k]} / \binom{r-1}{k}$ and $0 \leq j \leq n-1$; then
\[ \sum_{k=0}^{r-1} M^{[k]} \binom{n{-}1{-}j \; + \; r{-}k{+}T{-}1}{r{-}k{+}T{-}1} j^k \leq M n^{r-1} \binom{n{-}1{-}j{+}T}{T}. \]
\end{lem}

\begin{pf}
For $k \leq r-1$, we have
\[
\binom{n{-}1{-}j{+}T}{T}^{-1} \binom{n{-}1{-}j \; + \; r{-}k{+}T{-}1}{r{-}k{+}T{-}1}
= \frac{(n-j+T)^{\rising{r-1-k}}}{(T+1)^{\rising{r-1-k}}}
\leq (n-j)^{r-1-k};
\]
thus
\begin{multline*}
\binom{n{-}1{-}j{+}T}{T}^{-1}
\sum_{k=0}^{r-1} M^{[k]} \binom{n{-}1{-}j + r{-}k{+}T{-}1}{r{-}k{+}T{-}1} j^k
\leq \sum_{k=0}^{r-1} M^{[k]} j^k (n-j)^{r-1-k} \\ 
\leq M n^{r-1},
\end{multline*}
establishing the lemma.
\end{pf}

With $M$ as in Lemma~\ref{lem:M}, choose $K > M/\alpha$. Let $N_2 \geq N_1$ be such that $M n^{r} \leq \alpha K Q(n)$ for $n \geq N_2$. Suppose that some sequence $(v_n)$ satisfies $v_n \geq \abs{u_n}$ for $0 \leq n \leq N_2$ and
\begin{equation} \label{eq:majrec}
v_n = \frac{1}{n} \sum_{j=0}^{n-1} K \binom{n{-}1{-}j+T}{T} \alpha^{n-j} v_j
\end{equation}
for all $n \geq 1$. 
Let $n \geq N_2$. Assuming $\abs{u_j} \leq v_j$ for all $j \leq n-1$, and using
\eqref{eq:absrec} and Lemma~\ref{lem:M}, we get
\[
\frac{M n^r}{\alpha K} \abs{u_n} \leq Q(n) \abs{u_n} \leq \sum_{j=0}^{n-1} M n^{r-1} \binom{n{-}1{-}j+T}{T} \alpha^{n-1-j} v_j = \frac{M n^r}{\alpha K} v_n,
\]
hence by induction $\abs{u_n} \leq v_n$ for all $n \in \N$. Now \eqref{eq:majrec} translates into
\begin{equation} \label{eq:majdeq}
v'(z) = \frac{\alpha K}{(1-\alpha z)^{T+1}} v(z),
\end{equation}
which admits the simple solutions~\eqref{eq:normalv} below.

Finally, we adjust the integration constant $A$ so as to ensure that $\abs{u_n} \leq v_n$ for $n < N_2$ (lines~\ref{algo:boundnormaldiffeq:A:start}--\ref{algo:boundnormaldiffeq:A:end}). If no specific solution of~\eqref{eq:theta} is given (i.e., if we drop the parameter $u_{;n}$ of Algorithm~\ref{algo:boundnormaldiffeq}) we still obtain a result valid up to some multiplicative constant by simply ignoring this last part. The result of this computation is summarized in the following.

\begin{prop} \label{prop:boundnormaldiffeq}
Let $D \in \Q[z]\langle\theta\rangle$, and let $u_{;n}$ be a function that computes truncated series expansions of a specific $u \in \ker D$ up to any order $n$. Let $P_\alpha \in \Q[z]$. Assume that $0$ is a regular point of $D$ and that the dominant singularities of $D$ are finite and of modulus at least $\delta(P_\alpha)$. Then \BoundNormalDiffeq{$D, P_\alpha, u_{;\cdot}$} (Algorithm~\ref{algo:boundnormaldiffeq}) returns~$T \in \N$, $K \in \N^\ast$, $A \in \Q_+$ such that
\begin{equation} \label{eq:normalv}
u(z) \maj v(z) =
\begin{cases}
\displaystyle \frac{A}{(1-\alpha z)^K} & \text{if } T=0 \\
\displaystyle A \exp \frac{K/T}{(1-\alpha z)^T} & \text{otherwise.}
\end{cases}
\end{equation}
\end{prop}

In addition to its modulus $\alpha$, Algorithm~4 actually preserves the irregularity $T$ of the dominant singularity of the differential equation, which is connected to the subexponential growth of the coefficient sequence.

\begin{rem}
Sometimes all we need is a simple majorant series satisfying the tightness property of Theorem~\ref{thm:tightness} for the solutions of a differential equation of the form~\eqref{eq:dfinite} at an \emph{ordinary} point. Instead of the results of this section, we may then apply the ``plain'' Cauchy-Kovalevskaya method outlined in §\ref{sec:CK} using a majorant equation of the form
\[ v^{(r)} = \frac M {(1-\alpha z)^N} \sum_{k=0}^{r-1} \binom{r-1}{k} N^{\rising{r-k}} \Bigl( \frac \alpha {1-\alpha z} \Bigr) ^ {r-k} \; v^{(k)}. \]
This gives the majorant series $v(z) = \exp \bigl(M/(1-\alpha z)^N\bigr)$. If additionally the dominant singularity is regular, we may instead use the Euler equation
\[ v^{(r)} = \sum_{k=0}^{r-1} \frac{M^{[k]}}{(1-\alpha z)^{r-k}} v^{(k)}, \]
yielding $v(z) = A/(1-\alpha z)^\lambda$ where $\alpha^r \lambda^{\rising r} - M^{[r-1]} \alpha^{r-1} \lambda^{\rising{r-1}} - \dots - M^{[0]} = 0$. In both cases suitable parameters $M$, resp. $M^{[k]}$ may be determined using Algorithm~\ref{algo:boundratpoly}.
\end{rem}


\section{Explicit Bounds} \label{sec:explicit}

\subsection{P-Recursive Sequences} \label{sec:coeffs}

\begin{algorithm2e}[t]
\caption{Bounds for general P-recursive sequences} \label{algo:boundrecdoit}
\Function{\BoundRec{$R = \sum_{k=0}^s b^{[k]}(n) S^k \in \Q[n]\langle S\rangle, [u_0, \dots, u_{s-1}] \in \Q[i]^s$}}{
    $R \gets R \cdot S^{-m}$ where $m = \min \{k \mathbin| p^{[k]} \neq 0 \}$
        \nllabel{algo:boundrecdoit:valuation}
    \;
    $(\kappa, P_\alpha) \gets \Asympt(R)$
    \;
    \Rem{Normalize and encode the subexponential part by a differential equation}\;
    $D \gets \NormalizedDiffeq(R, \kappa)$
        \nllabel{algo:boundrecdoit:rectodiffeq}
    \;
    \Rem{Bound the solutions of the differential equation}\;
    define a function $\tilde u_{;\cdot}$ that ``unrolls'' the recurrence relation $R \cdot u = 0$ starting from $u_0, \dots, u_{s-1}$ to compute $\tilde u_{;n} = \sum_{k=0}^n q^{-pk/q} \Gamma(k/q+1)^{-p} u_k z^k$ (where $p/q=\kappa$) for any $n \in \N$
    \;
    $(T, K, A) \gets \BoundNormalDiffeq(D, P_\alpha, \tilde u_{;\cdot})$
    \;
    return $(\kappa, T, P_\alpha, K, A)$
}
\end{algorithm2e}

At this point, we are able to bound $u_n$ by a sequence $v_n$ given by its generating series $v(z) = \mathcal L'_{p,q} \tilde v(z)$, where $\tilde v$ is an explicit series satisfying a differential equation of the first order, and we have denoted 
\[ \mathcal L'_{p,q} v(z) = \sum_{n=0}^{\infty} \frac{v_n}{\psi_n} z^n. \]
(Note that series whose coefficients satisfy \emph{recurrence relations} of the first order, that is, hypergeometric series, cannot serve as asymptotically tight bounds for normalized D-finite functions because the range of asymptotic behaviours that their coefficient sequences assume is not wide enough: their ``subexponential'' asymptotic growth is always polynomial.)

\begin{prop} \label{prop:boundrec}
Given as input a nonsingular reversible recurrence operator $R \in \Q[n]\langle S \rangle$ along with initial values $u_0, \dots, u_{s-1} \in \Q[i]$ defining a solution $(u_n) \in \Q[i]^\N$ of $R·u=0$, the function $\BoundRec$ (Algorithm~\ref{algo:boundrecdoit}) computes $p/q \in \Q$, $P_\alpha \in \Q[z]$, $T \in \N$ and $K, A \in \R_+$ such that
\begin{equation}\label{nicebound}
\forall n \in \N, \quad \abs{u_n} \leq v_n = q^{\frac p q n} \, \Gamma\Bigl(\frac n q + 1\Bigr)^p \, \tilde v_n
\end{equation}
where $\tilde v_n$ is defined as in~\eqref{eq:normalv}. Additionally, for generic $(u_0, \dots, u_{s-1})$,
\[
\limsup_{n\to\infty} \left| \frac{u_n}{v_n} \right|^{1/n} = 1.
\]
\end{prop}

Allowing initial conditions in $\Q[i]$ rather than $\Q$ is convenient in view of some applications to numerical computations with D-finite functions (§\ref{sec:applications}).

\begin{pf}
This follows from combining the statements of Propositions~\ref{prop:tightness}, \ref{prop:diffeqtorec} and \ref{prop:boundnormaldiffeq}. Recall that we have chosen $\psi_n = q^{-\frac p q n} \Gamma(n/q+1)^{-p}$. After Line~\ref{algo:boundrecdoit:valuation} of Algorithm~\ref{algo:boundrecdoit}, the operator $R$ satisfies the hypotheses of Proposition~\ref{prop:diffeqtorec}. Hence the operator $D$ computed on Line~\ref{algo:boundrecdoit:rectodiffeq} cancels $\tilde u(z) = \sum_{n=0}^{\infty} \psi_n u_n z^n$, and the function $\tilde u_{;\cdot}$ defined on the next line does indeed compute truncations of this series. By Proposition~\ref{prop:boundnormaldiffeq} it follows that $\tilde u \maj \tilde v$ and, multiplying the coefficients by $\psi_n^{-1}$, that $u \maj v$. Finally, for generic initial values,
\[
\limsup_{n\to\infty} \left| \frac{u_n}{v_n} \right|^{1/n}
= \limsup_{n\to\infty} \left| \frac{u_n}{n!^\kappa \alpha^{n+o(1)} n^{O(1)}} \right|^{1/n}
= 1
\]
by Proposition~\ref{prop:tightness}.
\end{pf}

Although this representation \eqref{nicebound} is satisfactory for many applications, more explicit expressions for the coefficients $v_n$ are sometimes desirable. If $T=0$, it is readily seen that
\begin{equation} \label{eq:vn0}
\tilde v_n = A \, \alpha^n \binom{n+K-1}{K-1}.
\end{equation}
For $T > 0$, the general coefficient $\tilde v_n$ still admits a rather complicated ``closed-form'' expression in terms of the general hypergeometric function~$F$~\citep[see][§5.5]{ConcreteMath}: one may check that
\begin{align*}
\tilde v_n
= A \alpha^n \sum_{k=0}^\infty \frac{1}{k!} \binom{Tk+n-1}{n} \left(\frac K T\right)^k
= A \, \alpha^n \, {}_{T}\!F_T\!
\left(
\begin{matrix}
\frac{n+T}{T} & \frac{n+T+1}{T} & \cdots & \frac{n+2T-1}{T} \\
\frac{T+1}{T} & \frac{T+2}{T} & \cdots & \frac{2T}{T} 
\end{matrix}
\Bigg|
\frac K T
\right).
\end{align*}

However, $\tilde v_n$ may in turn be bounded by much simpler expressions without losing the asymptotic tightness (in the sense of Theorem~\ref{thm:tightness}) using a simple version of the saddle point method~\citep[see, e.g.,][§4.3]{AnalyticCombinatorics}. Since $\tilde v \in \R_+[[z]]$, for any $t \in (0;1/\alpha)$, we have $\tilde v_n \leq \tilde v(t)/t^n$. For fixed $n$, the right-hand side is minimal for the unique $t_n\in (0;1)$ such that $K \alpha t_n = n (1 - \alpha t_n)^{T+1}$. Asymptotically, $t_n$ satisfies
$1 - \alpha t_n \sim (K/n)^{1/(T+1)}$
as $n \to \infty$. This approximation suits our purposes well: indeed, we set 
\begin{equation} \label{eq:rn}
r_n = \frac1\alpha \left(1-\Big(\frac{K}{n+K+1}\Big)^{\frac 1 {T+1}} \right).
\end{equation}
(The term $K+1$ in the denominator does not change the asymptotic behaviour and is such that $r_n\in(0;1/\alpha)$.)
For $T > 0$, we obtain (with $A=1$)
\begin{equation} \label{eq:asymptvnirreg}
\begin{split}
\tilde v_n \leq \frac{\tilde v(r_n)}{r_n^n} =
\alpha^n
\left(1-\Big(\frac{K}{n+K+1}\Big)^{\frac 1 {T+1}} \right)^{-n}
\exp\left( \frac K T \left(\frac{n+K+1}{K}\right)^{\frac T {T+1}} \right) \\
= \alpha^n \exp O(n^{T/(T+1)}),
\end{split}
\end{equation}
and similarly
\begin{equation} \label{eq:asymptvnreg}
\tilde v_n \leq \alpha^n \Bigl(\frac{n+K+1}{K}\Bigr)^K  \Bigl(1-\frac{K}{n+K+1}\Bigr)^{-n} = \alpha^n n^{O(1)}
\end{equation}
if $T=0$.

Going back to $v_n$ itself, \eqref{eq:asymptvnirreg} and~\eqref{eq:asymptvnreg} extend to bounds of the form~\eqref{eq:thm}, that make the asymptotic behaviour $u_n = n!^\kappa \, \alpha^n \, e^{o(n)}$ apparent, by means of the following relation between $\psi_n$ and $n!^\kappa$.

\begin{lem} \label{lem:psivsfacto}
For $q \in \N\setminus\{0\}$ and $n \geq 3q/2$,
\[
\frac{1}{\psi_n}
= \Gamma(n/q+1)^p q^{p/q \; n}
\leq 
\begin{cases}
(2\pi)^{p/q} \, (n/q+1)^p \, n!^{p/q}, & p > 0 \\
n^{-p/q} \, n!^{p/q}, & p < 0.
\end{cases}
\]
\end{lem}

\begin{pf}
Since $\Gamma(x)$ is increasing for $x \geq 3/2$, 
\[
\prod_{k=0}^{q-1} \Gamma(n/q+k/q) \leq \Gamma(n/q+1)^q \leq \prod_{k=0}^{q-1} \Gamma(n/q+k/q+1).
\]
By Gauß' multiplication theorem \citep[see][Formula~6.1.20]{AbramowitzStegun1972}
\[ \Gamma(q z) = (2\pi)^{(1-q)/2} q^{q z - 1/2} \prod_{k=0}^{q-1} \Gamma\Bigl(z+\frac{k}{q}\Bigr) \qquad (z \in \CC), \]
this implies that
\[
\frac{(2\pi)^{(q-1)/2}}{n q^{-1/2}}
\leq
\frac{q^n \Gamma(n/q+1)^q}{\Gamma(n+1)}
\leq
\frac{(2\pi)^{(q-1)/2} (n+1)^{\rising{q-1}}}
     {q^{q-1/2}}
\]
and the result follows by raising either inequality to the power of $p/q$ depending on the sign of $p$.
\end{pf}

This concludes the proof of Theorem~\ref{thm:tightness}.

\begin{rem}
If we content ourselves with computing a numerical bound for one coefficient (or one tail, see next section) of a D-finite power series---that is, a bound for fixed $n$, as opposed to a formula giving a bound as a function of $n$---then majorant series with the same radius of convergence as the coefficients of the equation (and thus the method of §\ref{sec:dfinite}) are not strictly necessary for the bound to become ultimately tight as $n$ approaches infinity. Consider for instance Equation~\eqref{eq:prec} in the case where $0$ is an ordinary point, and assume $\nu > \alpha$ with the notations of §\ref{sec:dfinite}. \Citet[§3.5]{vdH2003} proves that if $p^{[k]}/p^{[r]} \maj M(\nu)/(1-\nu z)$ for $k=0, \dots, r-1$, then
\[ u(z) \maj \frac{C}{(1-\nu z)^{\ceil{(M(\nu)+1)/\nu}}} \]
where $C$ does not depend on $\nu$. Also assume that the majorizing procedure for rational functions used to compute $M(\nu)$ is tight enough to ensure that $M(\nu) = O \bigl(n^d (\alpha/\nu)^n \bigr)$ (as is Algorithm~\ref{algo:boundratpoly}, with $d = \max_{k=0}^{r-1} m_k$). In a manner somewhat reminiscent of the saddle-point method, we then choose, say, $\nu = \nu_n = (1+1/n^{1/(2d)}) \alpha$, hence getting
\[ \abs{u_n} \leq v_n = \alpha^{n+n^{1-1/(2d)}}. \]
This suggests that it is sensible to take $\nu = (1+1/n^{\Theta(1/d)}) \alpha$ in the algorithms of \citet{vdH2001, vdH2003}.
\end{rem}

\subsection{Tails of Power Series} \label{sec:tails}

In Examples~\ref{ex:intro}(\ref{item:exchudnovsky}) and~(\ref{item:Si}), the sequence for which we compute an upper bound is the tail $t_n = u_{n;}(1)$ of a convergent series whose coefficients $u_n$ are given by a linear recurrence relation of the form~\eqref{eq:prec}.
In such a case, the sequence $t_n$ is also P-recursive, but its initial values are unknown---if we have in mind the evaluation of the sum of the series, these initial values are precisely what we are after. However, if $u(z) \maj v(z)$, the general properties of majorant series (§\ref{sec:majorants}) ensure that $\abs{u_{n;}(1)} \leq v_{n;}(1)$. To avoid repeated majorant computations when working with D-finite power series, notably in the context of numerical analytic continuation (see~§\ref{sec:num}), we  actually consider the slightly more general problem of bounding the tails $u^{(j)}_{n;}(z)$ of the $j$-th derivative of $u$ at any point $z$ such that $\abs{z} < \delta(p^{[r]})$, where $p^{[r]}$ is the leading term of a differential equation with polynomial coefficients annihilating $u(z)$.

We assume once again that we have computed $\kappa = p/q$ and $\tilde v$ such that $u(z) \maj v(z) = \mathcal L'_{p,q} \tilde v(z)$ (with $p \leq 0$, so that the radius of convergence of $v$ is positive) using the algorithms of §\ref{sec:asympt} and §\ref{sec:majorants}. The letters $\alpha$, $T$, $K$ denote the parameters of $\tilde v$ appearing in \eqref{eq:normalv}.
The formalism of majorant series proves handy here, as we have $\mathopen| u^{(j)}_{n;}(z) \mathclose| \leq v^{(j)}_{n;}(\abs{z})$ by Lemma~\ref{lem:majorants}. Notice that if $p < 0$, the point $z$ lies within the disk of convergence of $v$ but not necessarily in that of $\tilde v$.

\begin{figure}
\begin{center}
\includegraphics[width=.5\textwidth]{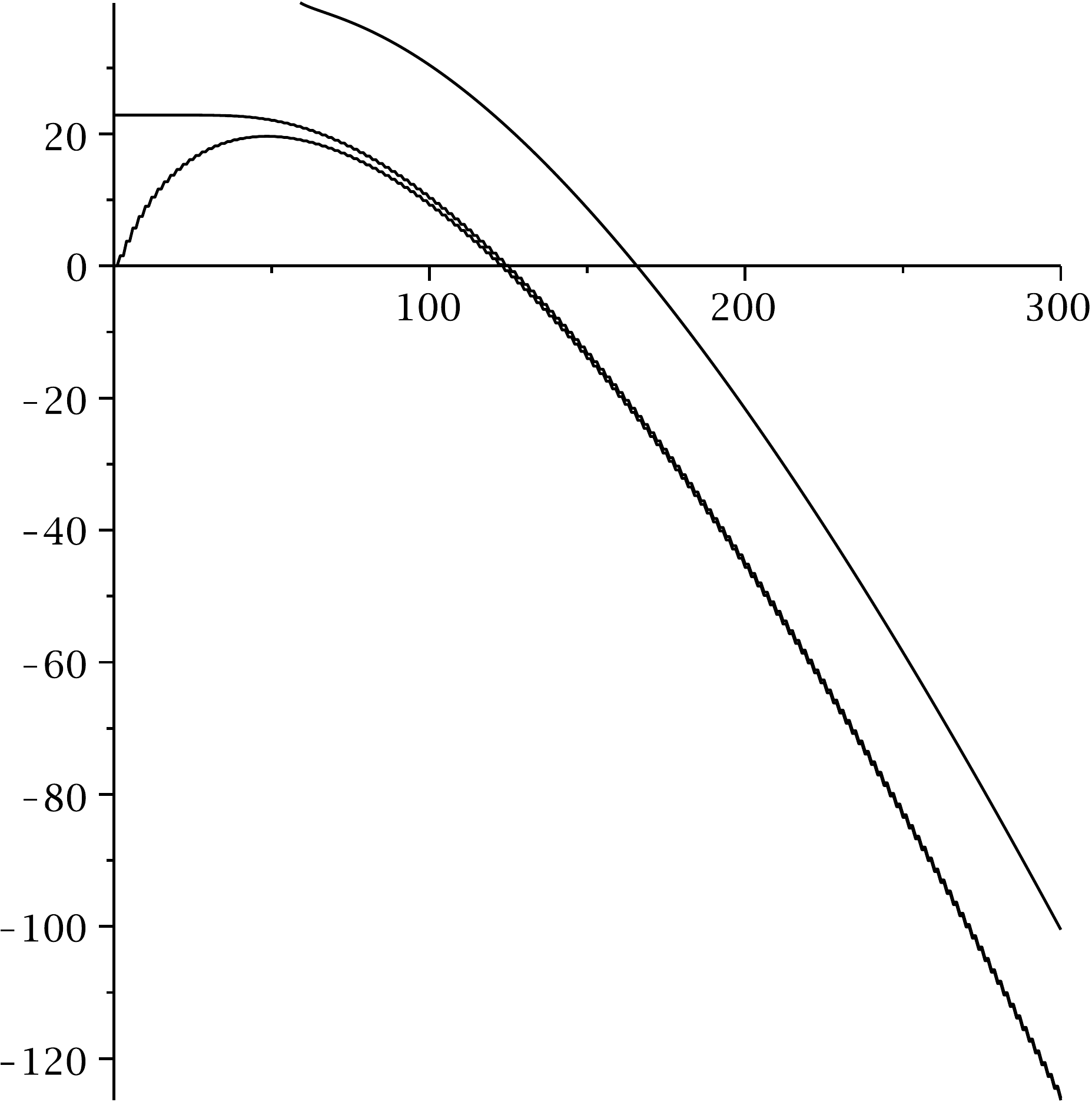}
\end{center}
\caption{From bottom to top, $\log(\erf_{n;}(5))$, $\log\abs{\erf_{n;}(5i)}$ and $\log (b(n))$ where $b(n)$ is the bound \eqref{eq:tails} with parameters computed by Algorithm~\ref{algo:boundrecdoit}.}
\label{fig:tailserf}
\end{figure}

\begin{prop}[Bound on $u_{n;}(z)$ for large $n$] \label{prop:tails}
With $z$ and $v$ as above, assume that
\begin{equation} \label{eq:condn}
n >
\begin{cases}
(1-\alpha \abs{z})^{-T-1} K, & \kappa=0 \\
\displaystyle{
(\alpha\abs{z}})^{-q/p}
\left(1 - \Bigl(\displaystyle \frac{K}{(\alpha \abs{z})^{-q/p}+K+1}\Bigr)^{\frac{1}{T+1}}
\right)^{q/p}, & \kappa < 0.
\end{cases}
\end{equation}
Then for all $j$, we have
\begin{equation} \label{eq:tails}
\abs{u_{n;}^{(j)}(z)}
\leq \frac{\tilde v^{(j)}(r_n)}{q^{-\frac p q n} \Gamma(\frac n q + 1)^{-p}} \Bigl(\frac{\abs{z}}{r_n}\Bigr)^n h\Bigl(\frac{\abs{z}}{r_n}\Bigr),
\end{equation}
where $r_n$ is given by \eqref{eq:rn} and
\[
h(x) = \frac{1}{1-x^q/(n+q)^{-p}} \sum_{u=0}^{q-1} x^u \qquad  \text{($=1/(1-x)$ for $\kappa=0$, i.e. $p/q=0/1$).}
\]
The bound~\eqref{eq:tails} is generically tight up to subexponential factors.
\end{prop}

Figure~\ref{fig:tailserf} illustrates the behaviour of this bound for entire functions, in the typical situation where the Taylor series at the origin ``starts converging'' only beyond a significant ``hump''.
Once again, the factor $n!^{p/q}$ in \eqref{eq:tails} can be brought out explicitly if desired using Lemma~\ref{lem:psivsfacto}. 

\begin{pf}
In the case $\kappa=0$, the condition~\eqref{eq:condn} ensures that $\abs{z} < r_n < \alpha^{-1}$. Using the relation $\tilde v_n = \psi_n v_n$ and the saddle-point bound $\tilde v_k \leq \tilde v(r_n)/r_n^k$ (notice the $n$), we obtain
\[
\abs{u_{n;}^{(j)}(z)}
\leq v_{n;}^{(j)}(\abs z)
\leq
  \frac{\tilde v^{(j)}(r_n)}{\psi_n} 
  \Bigl(\frac{\abs{z}}{r_n}\Bigr)^n
  \sum_{k=0}^\infty
    \frac{\psi_n}{\psi_{n+k}} \Bigl(\frac{\abs{z}}{r_n}\Bigr)^{k}.
\]
This proves~\eqref{eq:tails} for $\kappa=0$. Now assume $p < 0$, and recall that in this case $\psi_n = q^{-p/q} \Gamma(n/q+1)^{-p}$ is increasing: hence
\begin{align*}
\sum_{k=0}^\infty \frac{\psi_n}{\psi_{n+k}} x^k
\leq \sum_{t=0}^\infty \sum_{u=0}^{q-1} \frac{\psi_n}{\psi_{n+tq}} x^{tq}
= \sum_{u=0}^{q-1} x^u \sum_{t=0}^\infty \frac{x^{tq}}{\big( (n+q)(n+2q)\dots(n+tq) \big)^{-p}} \leq h(x)
\end{align*}
for $n \geq x^{-q/p}$. But this last condition follows from~\eqref{eq:condn} since
\[
\Bigl(\frac{\abs{z}}{r_n}\Bigr)^{-q/p}
<
(\alpha\abs{z})^{-q/p}
\left(1 - \Bigl(\displaystyle \frac{K}{(\alpha \abs{z})^{-q/p}+K+1}\Bigr)^{\frac{1}{T+1}}
\right)^{q/p}
\]
as soon as $n > (\alpha \abs{z})^{-q/p}$, itself implied by~\eqref{eq:condn}.

The estimates~\eqref{eq:asymptvnirreg},~\eqref{eq:asymptvnreg} still hold, hence the tightness of the bound.
\end{pf}

Bounds on $u_{n;}(z)$ are sometimes useful also when the condition~\eqref{eq:condn} fails to be satisfied, especially for $n=0$. Simple bounds independent on $n$ give good results.

\begin{prop}[Bound on $u_{n;}(z)$ for small $n$]
For all $n \in \N$ and $0 < r < \alpha^{-1}$,
\begin{equation}
\abs{u_{n;}^{(j)}(z)} \leq
\begin{cases}
v^{(j)}(\abs{z}) & \kappa = 0 \\
\displaystyle{
  v^{(j)}(r) \;
  \exp \Bigl{(} -\frac p q \Bigl{(}\frac{\abs{z}}{r} \Bigr{)}^{-q/p} \Bigr{)} \;
  \sum_{u=0}^{q-1} \Bigl{(} \frac{\abs{z}}{r} \Bigr{)}^u
}
& \kappa < 0.
\end{cases}
\end{equation}
\end{prop}

\begin{pf}
The proof is similar to that of Proposition~\ref{prop:tails}.
For $\kappa=0$ the result is obvious. Assuming $\kappa<0$, it holds for all $x > 0$  that
\[
\sum_{k=n}^\infty \frac{x^k}{\psi_k} \leq
\sum_{u=0}^{q-1} x^u \sum_{t=\floor{n/q}}^\infty \frac{x^{qt}}{\psi_{qt}} \leq
\sum_{u=0}^{q-1} x^u \sum_{t=\floor{n/q}}^\infty \frac{ (-\frac pq x^{-q/p})^{-pt} }{(-pt)!} 
\]
since $\psi_{qt} = q^{-pt} t!^{-p} \geq (-q/p)^{-pt} (-pt)!$ ($t \in \N$); whence
\[ 
\abs{u_{n;}^{(j)}(z)} \leq \tilde v^{(j)}(r) \sum_{k=n}^{\infty} \frac1{\psi_k}{\Bigl( \frac{\abs z}{r} \Bigr)^k} \leq v^{(j)}(r) \;
  \exp \Bigl{(} -\frac p q \Bigl{(}\frac{\abs{z}}{r} \Bigr{)}^{-q/p} \Bigr{)} \;
  \sum_{u=0}^{q-1} \Bigl{(} \frac{\abs{z}}{r} \Bigr{)}^u.
  \qedhere
\]
\end{pf}

In the important case where $\kappa = T = 0$ and $K \in \N$, the $v_{n;}(z)$ actually admit closed-form expressions of the form $(\alpha z)^n p(n)$, where $p \in \Q(\alpha z)[n]$. Indeed, starting from~\eqref{eq:normalv} and writing (for fixed $K$) $(n+k+1)^{\rising{K-1}} = \sum_{i=1}^K c^{[i]}(n) (k+1)^{\rising{i-1}}$, we get
\[
\left( \frac1{(1-\alpha z)^K} \right)_{\!n;}
= \frac{(\alpha z)^n}{(K-1)!} \sum_{k=0}^\infty (n+K+1)^{\rising{K-1}} (\alpha z)^k
= \frac{(\alpha z)^n}{(K-1)!} \sum_{i=1}^K \frac{(i-1)!}{(1-\alpha z)^i} c^{[i]}(n).
\]
This is the kind of formula that appears in Example~\ref{ex:intro}(\ref{item:exchudnovsky}). Such bounds are easier to read than~\eqref{eq:tails}, but they are numerically unstable due to cancellations. In a system providing numerical routines for hypergeometric functions, one can use the alternative expression
\[ \left( \frac1{(1-\alpha z)^K} \right)_{\!n;} = (\alpha z)^n \, \binom{n+K-1}{K-1} \,
{}_2\!F_1 \bigg(\begin{matrix}1 \quad n+K \\ n+1 \end{matrix} \bigg| \alpha z \bigg) \]
which does not suffer from this shortcoming.

Finally, note that it might be worthwhile looking for refined bounds in applications where $T$ is large and $\abs{z} \simeq \alpha^{-1}$, since~\eqref{eq:tails} becomes tight only for very large $n$ in this case. Similar issues exist when $K$ is too large; they may be mitigated by modifying Algorithm~\ref{algo:boundratpoly} to compute bounds of the form $p(z)+M/(1-\alpha z)^m, p \in \Q_+[z]$, which allows for a tighter choice of $K$.

\section{Applications and Experiments} \label{sec:applications}

\subsection{Implementation}

We have implemented the algorithms described in this article (with slight variations) in the computer algebra system Maple. Our implementation is part of a submodule called NumGfun of the Maple package gfun%
\footnote{\url{http://algo.inria.fr/libraries/papers/gfun.html}}%
, but the code computing bounds is largely self-contained. It provides routines that compute majorant series for rational polynomials (following~§\ref{sec:rat}) and D-finite functions (§\ref{sec:dfinite}, §\ref{sec:coeffs}), and symbolic bounds for P-recursive sequences specified either using recurrence relations (§\ref{sec:coeffs}) or as tails of D-finite series (§\ref{sec:tails}). All examples of this article were computed using this implementation%
\footnote{To be precise, using gfun~v.~3.48 under Maple~13.}.

It is also used by the Dynamic Dictionary of Mathematical Functions%
\footnote{\url{http://ddmf.msr-inria.inria.fr/}}%
, an interactive web-based handbook of D-finite functions currently under development. All contents of the Dictionary are automatically generated from a compact description of each function (basically, a differential equation and initial values) using a mix of symbolic computation algorithms and document templates. The webpages the system produces are interactive in that they allow the user to trigger more computations, typically by asking for ``more terms'' in an asymptotic expansion. This is a situation where being able to display human-readable formulae rather than merely computing numerical bounds represents a significant benefit. Code based on this article provides majorant series for the Taylor expansions of the functions, truncation orders for these expansions to reach a given accuracy over a given disk, and symbolic bounds for their tails involving the truncation order.

\subsection{Application to the Numerical Evaluation of D-Finite Functions} \label{sec:num}

Guaranteed numerical computation with entire classes of functions usually involves the \emph{automatic} computation of error bounds relating approximations, e.g., by truncated power series, to the functions they approximate. Elementary results from real and complex analysis commonly used to compute such error bounds include the alternating series criterion, Cauchy's integral formula, and several variants of Taylor's theorem. Karatsuba describes algorithms with error bounds for the evaluation of various special functions, including the hypergeometric function ${}_2F_1$ \citep[see][and the references therein]{Karatsuba1999}. \citet{DuYap2005} provide bounds for the tails of the general hypergeometric series, where the parameters are allowed to vary, based on a detailed analysis of the variations of the coefficient sequence. For the more general case of D-finite functions, another \emph{ad hoc} method is given by \citet{vdH1999}. In a different context, \cite{Neher2003} uses Cauchy's estimate and complex interval arithmetic to bound the coefficients and tails of series expansions of arbitrary ``explicit enough'' analytic functions. This method is implemented in ACETAF \citep{EbleNeher2003}.

A further classical tool is the Cauchy-Kovalevskaya majorant series method discussed in~§\ref{sec:CK}. This idea is exploited by \citet[§2.4]{vdH2001} to bound the tails of power series expansions of D-finite functions in the neighbourhood of an ordinary point of the equation, and later again in a much more general setting covering a wide range of functional equations~\citep{vdH2003}. This is the approach we rely on in this article: indeed, the algorithm we described in §\ref{sec:dfinite} may actually be seen as a refinement of those suggested in §3.5 and §5.2 of the latter article. The main originality of our approach is the asymptotic tightness of the bounds.

Finally, it should be noted that in the context of numerical evaluation, instead of using \emph{a priori} bounds, it is often easier to compute successive error bounds in parallel to successive approximations of the result, until the desired accuracy is reached. The computation of validated numerical enclosures of solutions of ODE, DAE and more general functional equations has been the subject of extensive literature since the sixties \citep[see][]{Rihm1994a} in the area of interval methods. Of special interest when working with power series is the integration of differential equations using Taylor models \cite[see][]{Hoefke2001,NeherJacksonNedialkov2007}. Taylor models are one among a fair number of different symbolic-numeric representations of functions used in interval arithmetic, several of which have a similar approach of bounds for solutions of functional equations: for more on Taylor models and their relation to other interval methods, see~\citep{MakBer2003, Neumaier2003}. Some of these methods were imported to computer algebra and revisited by \citet{vdH2007b} in the context of rigorous effective complex analysis.

In a nutshell, the common idea is to write the (differential, say) equation at hand in fixed-point form $u = \Phi(u)$, where $\Phi$ is an integral operator, and to consider the action of $\Phi$ on truncated power series augmented with error bounds, using rules such as
\[
\int^x (a_0 + a_1 t + a_2 t^2 + [\alpha, \beta]) dt \subseteq \int^x (a_0+a_1t) dt + B\!\Big(a_3 \frac{x^3}{3}\Big) + [\alpha, \beta] \cdot B(x).
\]
Here $B(p)$ is an interval containing the range of $p(x)$ obtained from the range of $x$. One then computes an approximate solution in the form of a Taylor expansion $p(x) = a_0 + \cdots + a_n x^n$ and iteratively searches for a tight interval $[\alpha,\beta]$ such that $\Phi(p+[\alpha,\beta]) \subset p+[\alpha,\beta]$, possibly narrowing the range of $x$ or increasing the expansion order $n$ as necessary. Under mild assumptions, the existence of such $p + [\alpha, \beta]$ implies that of an actual solution $u \in p + [\alpha, \beta]$ of $\Phi(u)=u$.

While this is reported to provide tight numerical enclosures at reasonable cost for computations at machine precision even in the case of nonlinear equations in many variables, we are not aware of any asymptotic tightness result of the kind we are interested in in this paper. In fact, it is not entirely clear to us under which conditions methods of this kind are guaranteed to produce arbitrarily tight enclosures. (Note however that \citet{vdH2007b} states initial results in this direction.) Neither do we know how to use them to bound tails of D-finite functions on their whole disk of convergence. 

And yet, D-finite functions may be evaluated to an absolute precision $10^{-n}$ in softly linear time $n (\log n)^{O(1)}$ by computing truncations of their Taylor series by binary splitting. Numerical analytic continuation based on this technique then allows to obtain values of these functions at any point of their Riemann surfaces~\citep[§5]{ChuChu1988}. Applications include the numerical computation of monodromy matrices of linear differential equations with polynomial coefficients. In this context, one benefit of the language of majorant series is that a single majorant encodes both bounds on the values and truncation orders for all elements of a basis of the local solutions of the differential equations as well as their derivatives---all of which are useful to control errors in the numerical analytic continuation process.

Excluding degenerated cases, the number of terms of the series to take into account is $\lambda n + o(n)$, where $\lambda$ depends on the location of the evaluation point relative to the singularities of the function, or $O(n/\log n)$ in the case of entire functions. The tightness result of Theorem~\ref{thm:tightness} translates into the fact that the number $N$ of terms that get computed is indeed of that order, while most existing methods for computing bounds of tails of D-finite series seem to ensure only $N = O(n)$. This in turn improves the complexity of the algorithm by a constant factor.

The subpackage of gfun mentioned above contains high-precision numerical evaluation and analytic continuation routines based on this strategy. They rely on the code computing bounds for accuracy control. These numerical evaluation facilities are exported to the DDMF.

\subsection{Experiments}

\begin{table}[p]
\newcommand\numcol{11}
\newcommand{\datai}{
& 	 \multicolumn{2}{l}{$\frac{1}{(1-z)^2}\,@\frac12$} & 	 \multicolumn{2}{l}{$\frac{\cos z}{1-z}\,@\frac12$} & 	 \multicolumn{2}{l}{$\frac{\cos z}{1-z^2}\,@\frac12$} & 	 \multicolumn{2}{l}{$\frac{\cos z}{(1-z)^2}\,@\frac12$} & 	 \multicolumn{2}{l}{$\frac{(z+1)^2 \cos z}{(z^3+z+1)^2}\,@\frac{1}{10}$}  \\
\hline
$10^{-10}$ 
& 	 40 & 40
& 	 46 & 34
& 	 54 & 33
& 	 54 & 39
& 	 24 & 12
 \\
$10^{-100}$ 
& 	 342 & 342
& 	 350 & 333
& 	 364 & 331
& 	 364 & 341
& 	 140 & 121
 \\
$10^{-1000}$ 
& 	 3336 & 3335
& 	 3346 & 3323
& 	 3366 & 3321
& 	 3366 & 3334
& 	 1232 & 1201
 \\
}
\newcommand{\dataibis}{
& 	 \multicolumn{2}{l}{$\frac{\arccot(z)}{(z^2-1)(z^2+5)}\,@\frac12$} & 	 \multicolumn{2}{l}{$\psi(1/2)\;$} & 	 \multicolumn{2}{l}{$\arctan \frac{1}{2}$} & 	 \multicolumn{2}{l}{$\arctan \frac{9}{10}$} & 	 \multicolumn{2}{l}{$\arctan \frac{99}{100}$}  \\
\hline
$10^{-10}$ 
& 	 64 & 27
& 	 40 & 23
& 	 44 & 28
& 	 336 & 164
& 	 4238 & 1496
 \\
$10^{-100}$ 
& 	 380 & 321
& 	 342 & 313
& 	 348 & 324
& 	 2338 & 2108
& 	 25210 & 21848
 \\
$10^{-1000}$ 
& 	 3392 & 3307
& 	 3336 & 3293
& 	 3344 & 3310
& 	 22050 & 21754
& 	 231844 & 227810
 \\
}
\newcommand{\dataii}{
& 	 \multicolumn{2}{l}{$\cos \frac{z}{1-z}\,@\frac13$} & 	 \multicolumn{2}{l}{$\sin \frac{z}{1-z}\,@\frac13$} & 	 \multicolumn{2}{l}{$\exp \frac{z}{(1-z)^2}\,@\frac12$} & 	 \multicolumn{2}{l}{$\exp \frac{z}{1-z^2}\,@\frac12$} & 	 \multicolumn{2}{l}{$\erf\left(\frac{1+z}{2z^2-1}\right)@\frac{1}{9}$}  \\
\hline
$10^{-10}$ 
& 	 48 & 25
& 	 46 & 24
& 	 118 & 79
& 	 68 & 42
& 	 28 & 12
 \\
$10^{-100}$ 
& 	 290 & 224
& 	 290 & 225
& 	 558 & 497
& 	 416 & 364
& 	 244 & 132
 \\
$10^{-1000}$ 
& 	 2416 & 2150
& 	 2416 & 2149
& 	 4154 & 4001
& 	 3566 & 3432
& 	 2384 & 1292
 \\
}
\newcommand{\dataiibis}{
& 	 \multicolumn{2}{l}{$\frac{\exp(1/(1-z))}{(1-z)}\,@\frac12$} & 	 \multicolumn{2}{l}{$\Bi\left(\frac{1}{1-z}\right)@\frac12$} & 	 \multicolumn{2}{l}{$\Ai\left(\frac{1}{1-z}\right)@\frac12$} & 	 \multicolumn{2}{l}{$\Ai\left(\frac{1}{1-z}\right)@\frac34$} & 	 \multicolumn{2}{l}{$\Ai\left(\frac{1}{1-z}\right)@\frac78$}  \\
\hline
$10^{-10}$ 
& 	 70 & 54
& 	 148 & 56
& 	 142 & 30
& 	 1558 & 77
& 	 23818 & 215
 \\
$10^{-100}$ 
& 	 418 & 387
& 	 664 & 416
& 	 660 & 345
& 	 3430 & 879
& 	 29258 & 2025
 \\
$10^{-1000}$ 
& 	 3568 & 3490
& 	 4700 & 3645
& 	 4694 & 3406
& 	 16284 & 8372
& 	 69594 & 18529
 \\
}
\newcommand{\dataiii}{
& 	 \multicolumn{2}{l}{$\Ai(4i+4)$} & 	 \multicolumn{2}{l}{$\Bi(4i+4)$} & 	 \multicolumn{2}{l}{$\Sinei(1)$} & 	 \multicolumn{2}{l}{$\cos(1)$} & 	 \multicolumn{2}{l}{$\sin(1)$}  \\
\hline
$10^{-10}$ 
& 	 92 & 59
& 	 92 & 59
& 	 16 & 12
& 	 18 & 13
& 	 18 & 14
 \\
$10^{-100}$ 
& 	 226 & 200
& 	 226 & 200
& 	 74 & 68
& 	 76 & 69
& 	 74 & 70
 \\
$10^{-1000}$ 
& 	 1054 & 1031
& 	 1054 & 1031
& 	 454 & 448
& 	 456 & 449
& 	 456 & 450
 \\
}
\newcommand{\dataiiibis}{
& 	 \multicolumn{2}{l}{$e^{-100}$} & 	 \multicolumn{2}{l}{$\erf^2(1)$} & 	 \multicolumn{2}{l}{$\erf(1)$} & 	 \multicolumn{2}{l}{$\erf(10)$} & 	 \multicolumn{2}{l}{$\erf(100)$}  \\
\hline
$10^{-10}$ 
& 	 298 & 291
& 	 60 & 33
& 	 36 & 24
& 	 628 & 574
& 	 54492 & 54388
 \\
$10^{-100}$ 
& 	 456 & 450
& 	 190 & 163
& 	 150 & 138
& 	 936 & 894
& 	 54904 & 54800
 \\
$10^{-1000}$ 
& 	 1406 & 1402
& 	 1036 & 1011
& 	 908 & 898
& 	 2828 & 2800
& 	 58870 & 58772
 \\
}
\begin{center}
\begin{tabular}{r*{5}{r@{/}l}}
\hline
\multicolumn{\numcol}{l}{Regular dominant singularity}\\
\hline
\datai
\hline
\dataibis
\hline \\
\hline
\multicolumn{\numcol}{l}{Finite irregular dominant singularity} \\
\hline
\dataii
\hline
\dataiibis
\hline \\
\hline
\multicolumn{\numcol}{l}{Dominant singularity at infinity} \\
\hline
\dataiii
\hline
\dataiiibis
\hline
\end{tabular}
\end{center}
\par\medskip
\caption{\label{table:benchs}Computed/minimal required number of terms of the Taylor expansion of a D-finite function to approximate this function to a given absolute precision. In this table, $\psi$ is the solution of the spheroidal wave equation $(1-z^2) \, \psi''(z) - 2(b-1)z \, \psi'(z) + (c-4qz^2) \psi(z)=0$ given by the choice of parameters and initial values $b=1/2$, $q=1/3$, $c=1$, $\psi(0)=1$, $\psi'(0)=0$; $\Ai$ and $\Bi$ denote the Airy functions; $\erf$ stands for the error function and $\Sinei$ for the integral sine.
}
\end{table}

In Table~\ref{table:benchs}, we report on experiments concerning the tightness of the bounds for truncating Taylor series expansions of a few common elementary and special functions. Each column label actually stands for a differential equation that annihilates the given function (with suitable initial values), and an evaluation point smaller in absolute value than the dominant singularity of the differential equation. Each internal cell shows the truncation order computed by the algorithm from this data for a specific accuracy requirement, and compares it to the minimal correct answer, computed by exhaustive search. For instance, the column ``$\erf(1)^2$'' corresponds to the evaluation at $z=1$ of the function $u(z) = \erf(z)^2$ represented as the unique solution of
\[
  (2+8z^2) \, u'(z) + 6z \, u''(z) + u'''(z), \qquad 
  u(0)=0, \quad u'(0)=0, \quad u''(0)=\frac8\pi.
\]
Using a majorant series for $u$, our algorithm determined that $\abs{u_{;190}(1) - u(1)} \leq 10^{-100}$, but it happens that only the first $163$ of these $190$ terms are really necessary. It can be seen that the bounds we compute do not stray too far from the optimal values.

We consider three cases, corresponding to the three main types of asymptotic behaviours that the coefficient sequence of a convergent D-finite series may exhibit, characterized (in generic cases) by the nature of the dominant singularities of the differential equation: regular singularities ($\kappa=0=T$ with the notations of the previous sections), irregular singularities at finite distance ($\kappa=0, T>0$), or at infinity ($\kappa < 0$). (Irregular singularities with $\kappa > 0$ correspond to divergent power series, and a differential equation whose only singularity is a regular singular point at infinity has only polynomial solutions. The examples of the second set all involve right composition by rational functions because it is unusual to study differential equations with more than two irregular singular points, and those are usually taken to be $\infty$ and $0$.)

For each of these, the last three columns illustrate how the truncation orders and the bounds vary as $\abs{z}$ approaches the radius of convergence of the series. Note that high-order Taylor expansions at $0$ are not the best way to compute numerical values of D-finite functions for such $z$: the growth of the truncation orders (both optimal and computed) can be got around by using several steps of analytic continuation along a broken-line path from $0$ to $z$ \citep[§4]{ChuChu1987}.

The example of $\Sinei(z)$ has an interesting feature: the origin is a regular singular point of the differential equation mentioned in Example~\ref{ex:intro}(\ref{item:Si}), but $\Sinei(z)$ may nevertheless be defined by simple initial values at origin, so that our algorithm applies without any adjustment.

Finally, here is a nontrivial ``non-generic'' example where our method fails to produce a tight bound.

\begin{exmp}
In his proof or the irrationality of $\zeta(3)$, \citet{Apery1979} introduces two sequences $(a_n)$ and $(b_n)$ such that $u_n = b_n - \zeta(3) a_n$ satisfies the (minimal-order) linear recurrence relation
\[ (n+2)^3 \, u_{n+2} = (2n+3)(17n^2+51n+39)\, u_{n+1} - (n+1)^3 \,u_{n}, \quad u_0=-\zeta(3), u_1=6-5\zeta(3). \]
Applied to this recurrence relation, Algorithm~\ref{algo:boundrecdoit} determines that 
\[
\abs{u_n} \leq 1.21 \, (n^2+3n+2) \, (17 + 12\sqrt{2})^{n} \qquad \text{(where $(17 + 12\sqrt{2}) \simeq 33.97$)}
\]
This bound is asymptotically tight for both $a_n$ and $b_n$, but the whole point of Apéry's proof is that $b_n - \zeta(3) a_n \to 0$ fast as $n \to \infty$.
\end{exmp}

\begin{ack}
We thank Moulay Barkatou, Nicolas Brisebarre, Sylvain Chevillard and Nicolas Le Roux for interesting discussions or comments on earlier versions of this work that have led to improvements, and an anonymous referee for spotting an error that made an important part of the reasoning hard to follow.
\end{ack}

\bibliographystyle{elsart-harv}

\end{document}